\documentclass[prl,aps,twocolumn,footinbib,preprintnumbers,superscriptaddress,preprintnumbers,balancelastpage,longbibliography]{revtex4-2}
\usepackage[utf8]{inputenc}
\usepackage[colorlinks=true,citecolor=blue,linkcolor=blue]{hyperref}
\usepackage[normalem]{ulem}
\usepackage{amsmath,amssymb, mathrsfs,xfrac}
\usepackage{epsfig}
\usepackage{graphicx}               
\usepackage{color}
\usepackage{slashed}
\usepackage{multirow}
\usepackage{placeins}
\usepackage{graphicx}
\usepackage[dvipsnames]{xcolor}
\usepackage{epstopdf}
\usepackage{soul}
\usepackage{tikz}
\usepackage[capitalise, english]{cleveref}
\usepackage{siunitx}
\usepackage{xspace}
\usepackage{booktabs}
\usetikzlibrary{trees}
\usepackage{graphicx}
\usepackage[export]{adjustbox}
\usetikzlibrary{decorations.pathmorphing}
\usetikzlibrary{decorations.markings}
\usepackage{aas_macros}
\usepackage{lettrine}
\usepackage{subfigure}
\usepackage{graphicx}
\usepackage{float}

\usepackage{placeins}
\usepackage{textgreek}
\usepackage{booktabs, tabularx, array}
\newcolumntype{Y}{>{\raggedright\arraybackslash}X}

\LettrineTextFont{\itshape}
\usepackage{comment}

\newcommand\myshade{80}

\colorlet{mylinkcolor}{ForestGreen}

\colorlet{mycitecolor}{Aquamarine}

\colorlet{myurlcolor}{violet}

\hypersetup{
  linkcolor  = mylinkcolor!\myshade!black,
  citecolor  = mycitecolor!\myshade!black,
  urlcolor   = myurlcolor!\myshade!black,
  colorlinks = true
}

\definecolor{jblue}{RGB}{20,50,100}
\definecolor{npurple}{RGB} {153, 51, 204}
\definecolor{wred}{RGB}{217,0,56}
\definecolor{white}{RGB}{255,255,255}

\definecolor{korange}{RGB}{235, 80,  43}
\definecolor{korange2}{RGB}{245, 100,  63}
\definecolor{kyelloworange}{RGB}{255, 210,  110}
\definecolor{kyelloworange2}{RGB}{240, 170,  90}
\definecolor{kred}{RGB}{204,  102, 153}
\definecolor{kpurple}{RGB}{153,  61, 190}
\definecolor{kpurplelight}{RGB}{213,  161, 230}

 \definecolor{tobycolour}{rgb}{.5,.0,.5}

\DeclareSIUnit\year{yr}
\DeclareSIUnit\pc{pc}
\DeclareSIUnit\ergs{ergs}
\DeclareSIUnit\msun{\ensuremath{M_\odot}}
\allowdisplaybreaks

\newcommand{\citephotoz}{Castellano2022, Finkelstein2022, Naidu2022b, Atek2023, Donnan2023, Harikane2023, Yan2023, Robertson2023, Adams2023, Bouwens2023b, Hainline2024}
\newcommand{\citespecz}{CurtisLake2023, DEugenio2024, Robertson2023, Carniani2024, Bakx2023, Castellano2024, Zavala2024, Fujimoto2023c, Wang2023b}

\makeatletter
\providecommand*{\diff}%
  {\command{\lmultau}{\ensuremath{L_\mu-L_\tau}\xspace}
\new@ifnextchar^{\DIfF}{\DIfF^{}}}
\def\DIfF^#1{%
  \mathop{\mathrm{\mathstrut d}}%
    \nolimits^{#1}\gobblespace}
\def\gobblespace{%
  \futurelet\diffarg\opspace}
\def\opspace{%
  \let\DiffSpace\!%
  \ifx\diffarg(%
    \let\DiffSpace\relax
  \else
    \ifx\diffarg[%
      \let\DiffSpace\relax
    \else
        \ifx\diffarg\{%
        \let\DiffSpace\relax
      \fi\fi\fi\DiffSpace}

\usepackage{tikz,xcolor,hyperref}

\definecolor{lime}{HTML}{A6CE39}
\DeclareRobustCommand{\orcidicon}{\hspace{-1mm}
	\begin{tikzpicture}
	\draw[lime, fill=lime] (0,0) 
	circle [radius=0.16] 
	node[white] {{\fontfamily{qag}\selectfont \tiny \,ID}};
	\draw[white, fill=white] (-0.0525,0.095) 
	circle [radius=0.007];
	\end{tikzpicture}
	\hspace{-3mm}
}

\foreach \x in {A, ..., Z}{\expandafter\xdef\csname orcid\x\endcsname{\noexpand\href{https://orcid.org/\csname orcidauthor\x\endcsname}
			{\noexpand\orcidicon}}
}

\usepackage{physics}


\newcommand{\mytitle}{Dark Secrets of Baryons:\,\,Illuminating Dark Matter-Baryon Interactions with JWST}

\begin{document}

	\title{\mytitle}
	
\author{Souradeep Das\orcidA{}}
\email{das.536@osu.edu}
\affiliation{Centre for High Energy Physics, Indian Institute of Science, C.\,V.\,Raman Avenue, Bengaluru 560012, India}
\affiliation{Department of Physics, Ohio State University, 191 West Woodruff Avenue, Columbus, OH 43210}

\author{Ranjini Mondol\orcidB{}}
\email{ranjinim@iisc.ac.in, ranjini12m@gmail.com,ranjini.mondol@saha.ac.in}
\affiliation{Centre for High Energy Physics, Indian Institute of Science, C.\,V.\,Raman Avenue, Bengaluru 560012, India}
\affiliation{Theory Division, Saha Institute of Nuclear Physics, 1/AF, Bidhannagar, Kolkata 700064, India}

\author{Abhijeet Singh\orcidC{}}
\email{abhijeets@iisc.ac.in}
\affiliation{Centre for High Energy Physics, Indian Institute of Science, C.\,V.\,Raman Avenue, Bengaluru 560012, India}

\author{Ranjan Laha\orcidD{}}
\email{ranjanlaha@iisc.ac.in}
\affiliation{Centre for High Energy Physics, Indian Institute of Science, C.\,V.\,Raman Avenue, Bengaluru 560012, India}

	\date{\today}
	
\begin{abstract}
The James Webb Space Telescope (JWST) has discovered bright galaxies at high redshifts ($z\approx 10-14$) and various galaxy candidates extending to even higher redshifts ($z\approx 15-30$).
Many astrophysical and beyond the Standard Model physics scenarios have been proposed to explain these observations. We investigate, {\it for the first time}, the implications of dark matter (DM) scattering with baryons (protons and electrons) in light of the JWST UV luminosity function (UVLF) observations. These interactions suppress structure formation on galactic scales, which may have an observable effect on the UVLF measurements at high redshifts. Using a recent galaxy formation model designed to explain high redshift observations, we obtain strong upper limits on DM-baryon scattering cross-sections and explore new regions of the parameter space. For DM-proton scattering with cross-section $\propto v^{-2}$ velocity dependence, we obtain the strongest limit for DM masses of $\sim$ 1 -- 500 MeV. For other cases that we study (DM-proton scattering cross-section $\propto v^{0},\,v^{-4}$ and DM-electron scattering cross-section $\propto v^{0},\,v^{-2},\,v^{-4}$, our limits are competitive with those obtained from other cosmological observables.
Our study highlights the potential of JWST observations as a novel and powerful probe of non-gravitational interactions of DM.
\end{abstract}
	
\maketitle
\footnotetext{The first two authors jointly led this research work.}

\maketitle
\section{Introduction}

In its first few years of operation, JWST has observed some of the earliest galaxies, which formed within $\sim 300$ million years after the Big Bang\,\cite{\citephotoz, 2025Ap&SS.370...85H, 2025arXiv250821708R, 2025arXiv250814972D, 2025ApJ...988..246M, 2025arXiv250708245T,2025arXiv250512505J,2025ApJ...985...80R,2025arXiv250511263N, 2025ApJ...983L..22K,2025arXiv250405893C, 2025arXiv250100984W, 2023MNRAS.518.4755A,Harikane2024a,Harikane2024b, 2025arXiv250706292W, Perez-Gonzalez:2025bqr, Yung:2025ttv, Fujimoto:2023fgh, 2025A&A...693A..50N, Davari:2023tam}. Using photometry, astronomers identified numerous galaxy candidates, many of which were later confirmed spectroscopically\,\cite{\citespecz,  2025arXiv250511263N, 2025ApJ...980..138H, 2024ApJ...976..193R}. While the findings of the JWST are  consistent with our current cosmological paradigm, observations have reported a surprisingly high abundance of bright galaxies at very early epochs ($z \gtrsim 10$). These galaxy abundances exceed expectations from pre-JWST models calibrated with \textit{Hubble Space Telescope} (HST) and \textit{Spitzer Space Telescope} data\,\cite{Boylan-Kolchin:2022kae, 2024ApJ...961...37C}. Several studies have explored the implications of these observations on galaxy formation models at very high redshifts\,\cite{2024A&A...690A.108L, 2025arXiv250804768S, 2025MNRAS.540.3350A,2024A&A...689A.244C, Dekel2023, 2010ApJ...710L.142F, 2016MNRAS.455..334T, 2024MNRAS.527.5929Y, 2024A&A...686A.138C,2024MNRAS.529.3563T, Menon2024,Sun2023a, Sun2023b, 2025ApJ...988L..10K, 2024AJ....168..113C, 2025ApJ...979..193C, 2024MNRAS.533.1111E, 2025A&A...697A..88L, 2023MNRAS.522.6236T, 2024ApJ...964..150D,2025NatAs...9..729H, 2025arXiv251004709F, 2025arXiv250919427S, 2025arXiv250919427S, 2025arXiv250723742F, 2024arXiv240300050K, Harikane2024b, 2025ScVar15595T, 2025ApJ97889H, 2024arXiv241207598N, 2024PDU4601600B,2024PhRvD.110j3540H,2024ApJ...975..285W, 2024ApJ...976L..16M,2024arXiv241011680D, 2024ApJ...974...27N,2024A&A...690A.108L, 2024AJ....168..113C, 2024PASJ...76..850I,2024MNRAS.532..149L,2024JCAP...07..078C,2024MNRAS.531.2615C, 2024PDU....4401496I, 2024A&A...684A.207F, 2024RAA....24d5001W,2024Natur.628..277G,2024ApJ...963...74W,2024MNRAS.529..628V,2023MNRAS.526.1324Q, Sun2023b,2023JCAP...10..012F,2023A&A...677L...4P, 2023ApJ...954L..48W, 2023MNRAS.524.3385G} and the impact of Beyond the Standard Model (BSM) physics on cosmological structure formation\,\cite{Klypin2021, Shen2024, Biagetti2023,Liu2022, Hutsi2023, Yuan2024, Dolgov2023, Parashari2023, Gong2023,Du:2024afd, 2025ApJ...993...17N, 2025A&A...702A.109U, 2025PhRvD.112f3560K, 2025PhRvD.111f3516D, 2024AstBu..79..535P, Menci:2024hop, 2024MNRAS.534.2848D, 2024PhRvD.110h3530T, 2024MNRAS.533.3923S, 2024MNRAS.533..860L, 2024PhRvD.110d3517W, 2024EPJP..139..711W,2024JCAP...07..072M,2024MNRAS.531.1021P,2024PhRvD.109l3002G,2024arXiv240615546C,2024ApJ...968...79L,2024PhRvD.109j3522P,2024A&A...685L...8C, 2024JCAP...05..097F, 2024arXiv240417803P, Hirano2015, 2024RAA....24a5009L, 2023SCPMA..6620403W, 2023JCAP...10..072A, 2023ApJ...953L...4P}.

Within the standard cosmological framework, $\sim 26.5$\% of the current energy density of the Universe is attributed to cold dark matter (CDM), whose existence has been established only through its gravitational influence on visible matter and radiation\,\cite{2022MNRAS.516.3556D, Planck:2018vyg, Massey:2010hh, 2022ApJ...934...43S, Mastropietro:2007kr, 2020Galax...8...37S, 2025MNRAS.542.2987S, 2025JCAP...01..021L, Cirelli2024}. However, the fundamental nature of DM remains unknown, and it may couple non-gravitationally to Standard Model (SM) particles\,\cite{Bertone:2016nfn, Cirelli2024, Strigari:2012acq, Slatyer:2017sev, Lin:2019uvt, Akita:2023yga}. Over the past few decades, researchers have sought to probe such interactions using a variety of astrophysical and cosmological observations, as well as terrestrial experiments. 

\begin{figure}[htb]
    \centering
    \includegraphics[width=1.0\columnwidth]{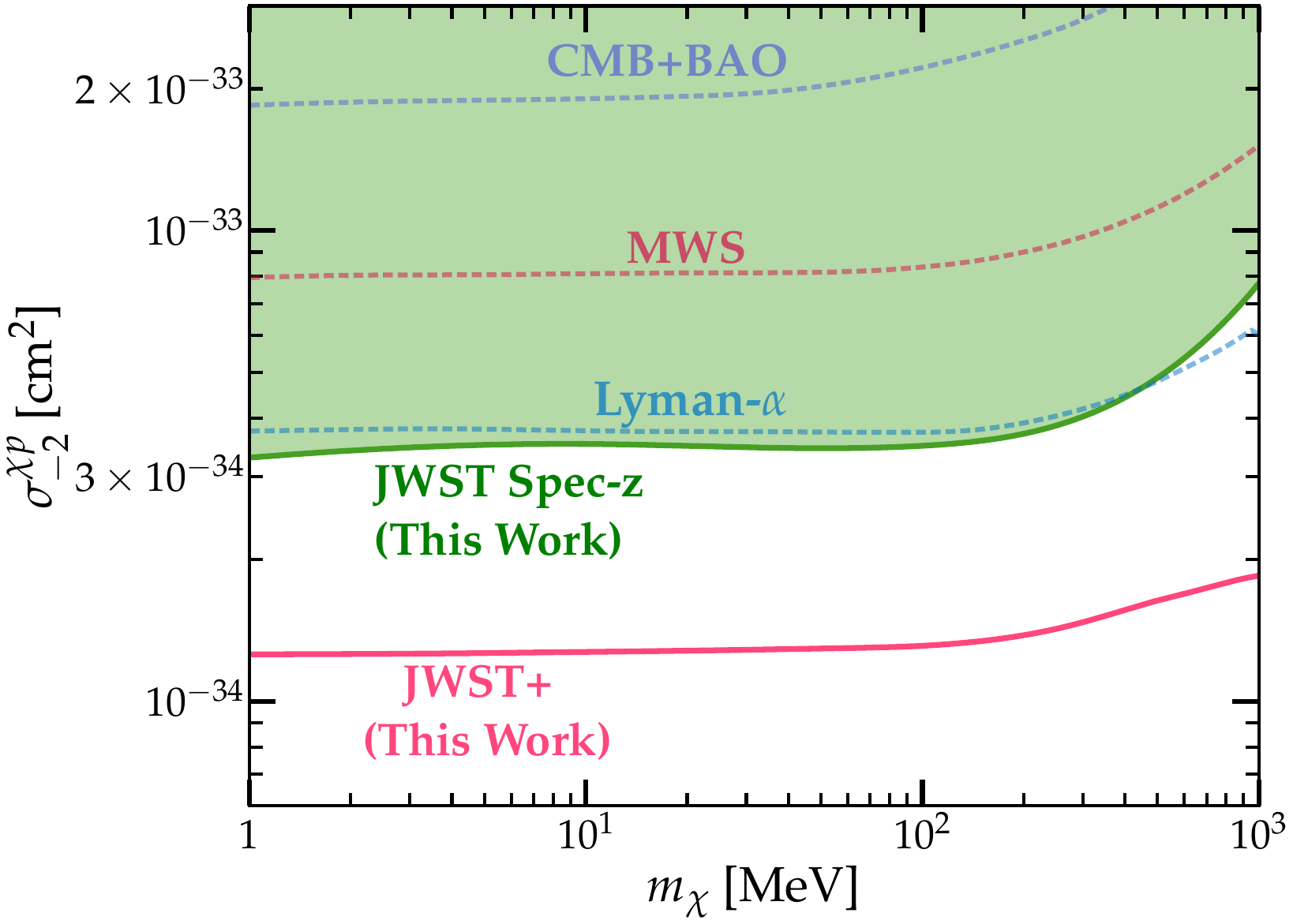}
    \caption{95\% C.L.\,\,upper limits on the normalization of the velocity-dependent \textit{electric-dipole-moment-like} ($\sigma\propto v^{-2}$) DM-proton elastic scattering cross-section as a function of the DM particle mass, $m_\chi$. The \textbf{green} shaded region is excluded by the ``JWST Spec-z" setup, while the \textbf{deep pink} line shows the upper limit on the normalization derived from the ``JWST+'' setup defined in the main text. Other dashed curves represent additional cosmological constraints -- Lyman-$\alpha$ forest (cyan), Milky Way satellite count (pink), and CMB+BAO (baryonic acoustic oscillations) (purple)\,\cite{Buen-Abad2022, Nguyen2021}.}
    \label{fig:n_minus_2_proton_Spec_z}
\end{figure}

Among the many possible types of interactions, DM–baryon (electrons or protons) elastic scattering has been extensively studied for a broad class of theoretical models\,\cite{Ali-Haimoud:2023pbi, Nguyen2021, Blanco:2023bgz, Buen-Abad2022, Gluscevic:2017ywp, Dvorkin2014, Maamari2021, Sigurdson:2004zp, Zhang:2024mmg}. Depending on the underlying particle-physics model, such interactions lead to the exchange of both heat and momentum between DM and baryons, resulting in shallower gravitational potential wells. These weakened potentials delay the gravitational collapse of matter, affecting the accumulation of DM and baryons, thus, suppressing structure formation\,\cite{Maamari2021, Nguyen2021, Ali-Haimoud:2023pbi, Dvorkin2014, Boddy2018, Boddy2018}. DM-baryon interactions have been examined through their signatures on the Cosmic Microwave Background (CMB)\,\cite{Ali-Haimoud:2021lka, Nguyen2021, Boddy2018}, the cosmological 21-cm signal\,\cite{Driskell2022, Rahimieh:2025fsb, Rahimieh:2025lbf, Barkana:2018lgd}, the Lyman-$\alpha$ forest\,\cite{Bose:2018juc, Buen-Abad2022}, the observed population of Milky Way satellite (MWS) galaxies\,\cite{Esteban2023, Buen-Abad2022, Maamari2021, Nadler:2019zrb}, and galaxy clustering data\,\cite{He:2025npy}. 

We investigate elastic scattering between DM and baryons using the JWST UV luminosity function (UVLF) data.\,\,For such interactions, the momentum-transfer cross section between DM and baryons, $\sigma^{\chi B}_{T,n}$, is parametrized as:
\begin{equation}\label{eq:v-dep}
    \sigma^{\chi B}_{T,n} = \sigma^{\chi B}_{n}v^{n},
\end{equation}
where $\sigma^{\chi B}_{n}$ is the normalization for the scattering cross-section, $B \in \{p, e^\pm\}$, is the target particle, $v$ is the relative velocity between DM and baryons, and $n$ is an exponent which may be positive or negative, depending on the underlying interaction model. We examine $n=0$, $n = -2$, and $n = -4$ cases, corresponding respectively to {\it contact}, \textit{electric-dipole-moment-like}, and \textit{Coulomb-like} interactions.

The UV luminosity of a galaxy and the mass of its host DM halo are correlated\,\cite{2022PhRvD.105d3518S}; thus UVLF measurements can serve as sensitive probes of these interactions. We adopt the modeling framework provided by the simulation suite \textsc{Thesan-Zoom}\footnote{\href{https://www.thesan-project.com/thesan-zoom/index.html}{https://www.thesan-project.com/thesan-zoom/index.html}} in order to connect the DM halo mass to galaxy UVLF\,\cite{2025arXiv250300106M, 2025arXiv250220437K, Shen:2025isu, 2025arXiv250303806Z, 2025arXiv250302927Z, 2025arXiv250304894M, 2025arXiv250505554W, 2025arXiv250708787M, 2025arXiv251013977P}.

\begin{figure*}
	\begin{center}
		\includegraphics[width=0.99\columnwidth]{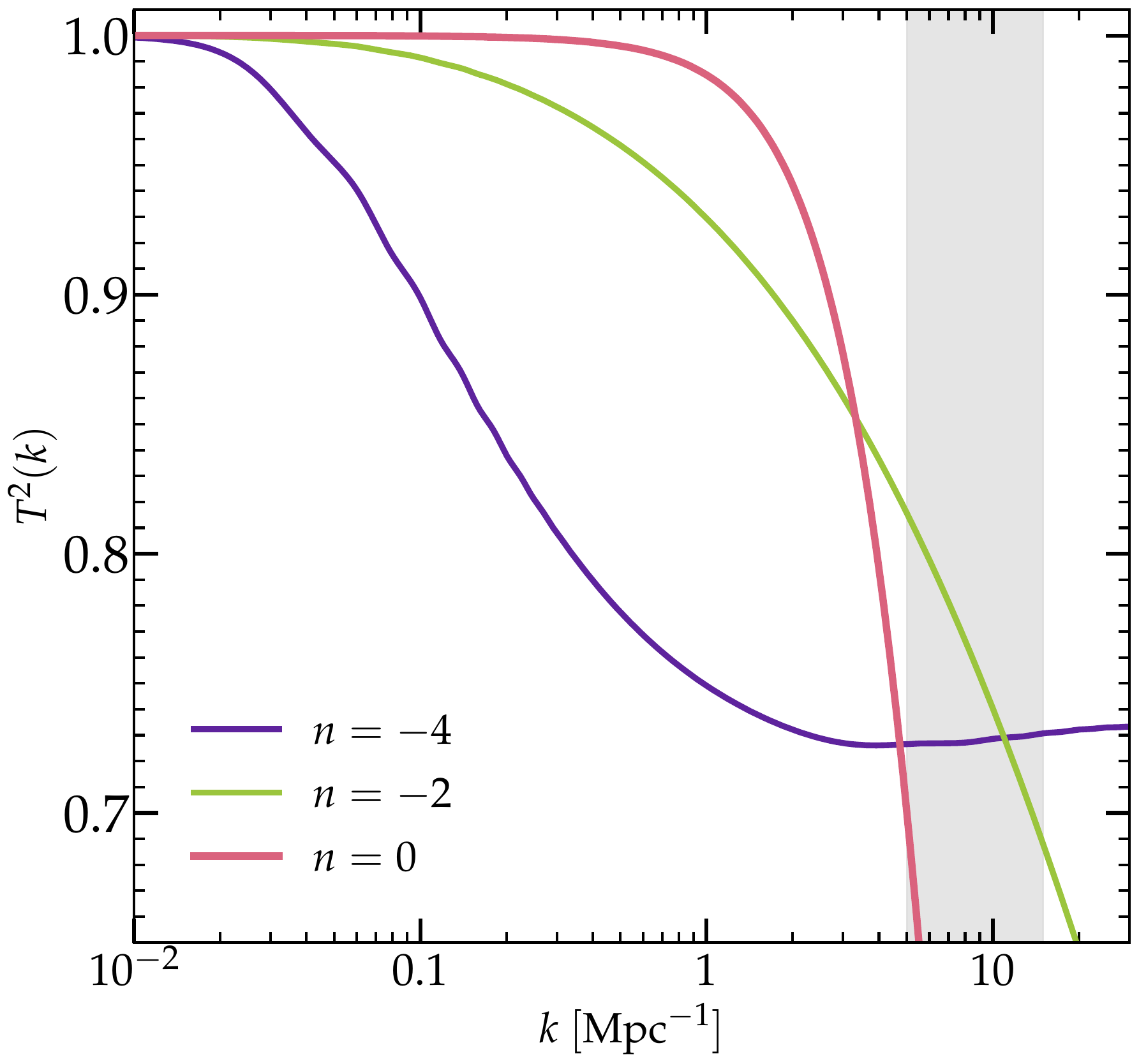}~~
	\includegraphics[width=0.99\columnwidth]{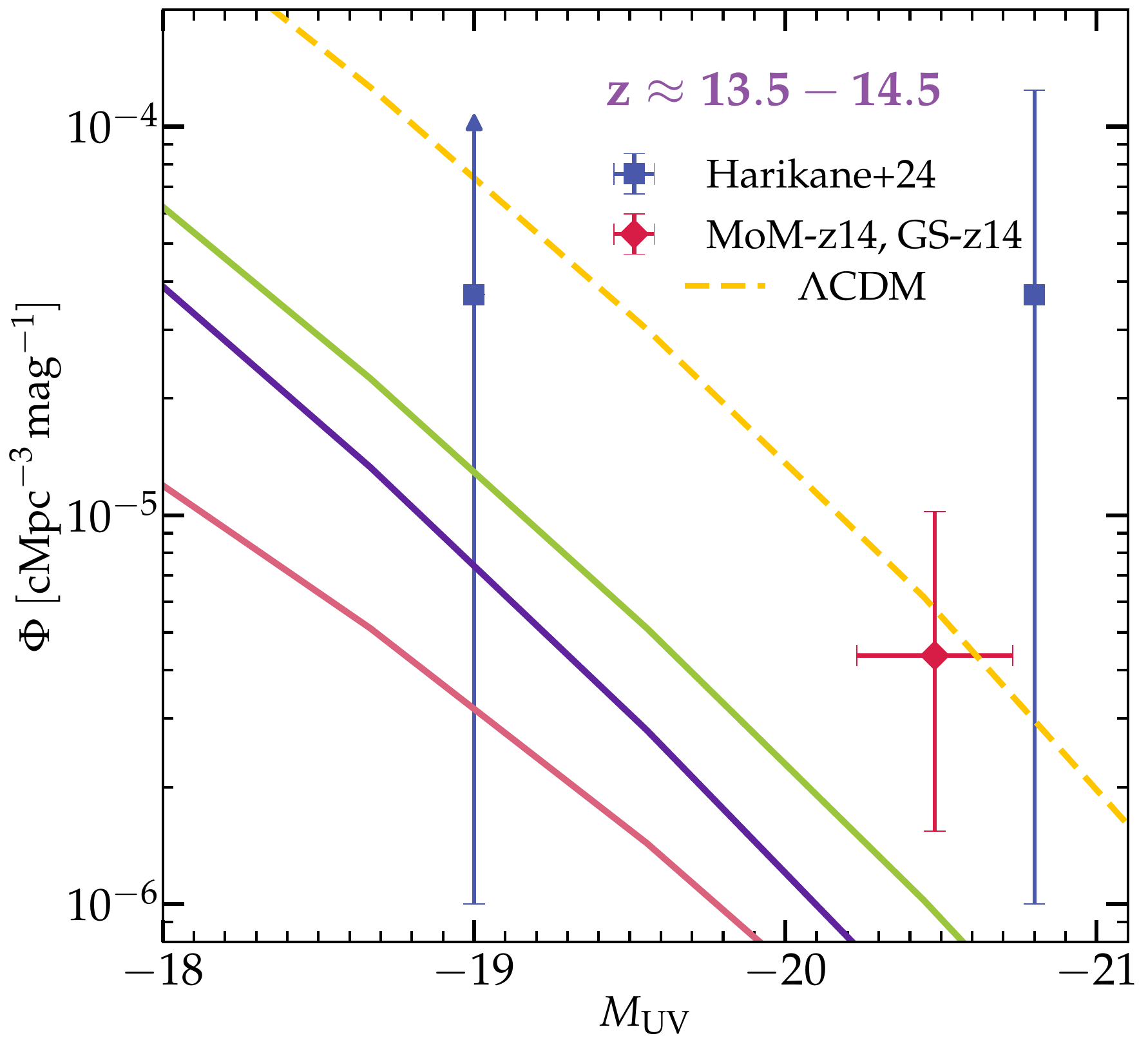}~~\\	
		\caption{
\textbf{\textit{Left:}} Comparison of the ratio $T^2(k) = P_{\rm IDM}(k)/P_{\rm \Lambda CDM}(k)$ of power spectrum of IDM to that of $\Lambda$CDM. We consider IDM cases with different values of $n$:\,\,$n = 0$ (\textbf{pink}), $n = -2$ (\textbf{green}), and $n = -4$ (\textbf{purple}), for DM of mass $m_\chi = 1~\rm{MeV}$, interacting with protons. 
Cross-section for each case corresponds to the 95\% C.L. upper limits obtained in this work: $\log_{10}(\sigma^{\chi p}_{n}/\mathrm{cm}^{2})$ for $n = 0,-2,-4$ cases are $-27.49$, $-33.48$, and $-40.33$, respectively using ``JWST Spec-z" analysis. Suppression of power spectrum at the characteristic scales (\textbf{gray shaded region}) is visible.
\textbf{\textit{Right:}} Comparison of the UVLFs at $z\approx 14$ for $\Lambda$CDM (\textbf{dashed yellow}) and IDM (same color code as \textit{left} panel) for the same IDM scenarios. Data points represent UVLF measurements at $z \approx 14$ as derived in Harikane et al.\,\cite{Harikane2024b} (\textbf{blue squares}) and from the MoM-z14 + GS-z14 galaxy observations\,\cite{2025arXiv250511263N} (\textbf{red diamond}). The choice of astrophysical parameters for $\Lambda$CDM correspond to the best-fit values from\,\cite{Shen:2025isu} with $\log_{10}({M_{c}/M_{\odot})} = 10.5$, while those for the interacting scenarios are fixed at the best-fit values obtained from our model by MCMC sampling.}
		\label{fig:uvlf-transfer-proton}
	\end{center}	
\end{figure*}

Fig.~\ref{fig:n_minus_2_proton_Spec_z} shows the 95\% confidence level (C.L.) upper limits on the normalization of the DM-proton interaction cross section, $\sigma^{\chi p}_{-2}$, for velocity-dependent ($n = -2$) interactions as a function of the interacting DM (IDM) mass, $m_{\chi}$.\,We use the JWST spectroscopic UVLF measurements of $\sim 40$ galaxies from refs.\,\cite{Harikane2024a, Harikane2024b, 2025arXiv250511263N} and the top-hat window function; this setup is referred to as ``JWST Spec-z". We find that the upper limits on $\sigma^{\chi p}_{-2}$ from this setup  (green solid line) provide the strongest constraint among existing cosmological observables. We also use a modified setup called ``JWST+". Here we include the JWST spectroscopic UVLF measurements from refs.\,\cite{Fujimoto:2023fgh, 2025A&A...693A..50N} together with photometric measurements at median redshifts $z \sim 17$ and $z \sim 25$ from \cite{2025arXiv250706292W, 2025arXiv250405893C, 2025ApJ...983L..22K, Perez-Gonzalez:2025bqr, 2025arXiv250100984W}, in addition to the previously used spectroscopic datasets; along with the smooth-$k$ window function. This is a more appropriate filter since the \textit{top-hat} filter function does not sufficiently damp the contributions at small scales and includes residual fluctuations\,\cite{Leo:2018odn}, overestimating the number of small halos. As expected, the inclusion of ultra-high-redshift data improves our constraints by a factor of $\sim 3$ (deep pink solid line), since higher redshifts correspond to larger $k$ values, making the data sensitive to smaller DM-baryon scattering cross-sections. For both setups, we use the Pantheon likelihood and impose Gaussian priors on $\omega_{b}$\,\cite{2009PhR...472....1I}, and on $A_{s}$ and $n_{s}$ from CMB measurements\,\cite{Planck:2018vyg}.  The suppression of the matter power spectrum can be quantified by the transfer function $T(k)$, whose square is the ratio of the matter power spectra for IDM and $\Lambda \mathrm{CDM}$. In fig.~\ref{fig:uvlf-transfer-proton}, $T^{2}(k)$ decreases gradually with increasing comoving wave number, $k$, resulting in a similar level of suppression across the $k$-scales probed by JWST, Lyman-$\alpha$, and MWS observations. Consequently, the limits from these probes are comparable. However, since JWST reports an overabundance of high-redshift structures, the suppression effect due to IDM is more incompatible with the JWST data, leading to the strongest constraints among all probes for sub-GeV masses of IDM.\,\,Additional constraints on individual DM masses for this case have been derived from other independent observational probes\,\cite{He:2025npy, Hooper:2022byl}. 

Since the ``JWST+'' setup includes photometric data as well, this result may be subject to change. For this reason, we derive constraints using this new setup only for the $n$ values where JWST already provides the strongest limits or is expected to do so as more data become available. We show our $n = -4$ constraints from this setup in \textit{Supplemental Material} \cite{suppmat}.
\nocite{An:2024nsw, Ma1995}
\nocite{Yin:2018yjn, CRESST:2019jnq, CRESST:2019axx, CRESST:2017ues, EDELWEISS:2019vjv, Ema:2018bih, Emken:2021lgc, Erickcek:2007jv, Mahdawi:2018euy, Essig:2012yx, Essig:2017kqs, XENON10:2011prx, XENON:2016jmt, DarkSide:2018ppu, SuperCDMS:2018mne, Crisler:2018gci, SENSEI:2019ibb, SENSEI:2020dpa, Fortin:2011hv, Essig:2011nj, Chu:2011be, Lesgourges2011, Press1974, Murray2013, Sheth:1999mn, Madau2014, 2012MNRAS.420.2318L, 2013MNRAS.428.1774B}

\section{DM-Baryon Interaction}\label{sec:DM-b-Interaction}

We study DM-baryon elastic scattering for two distinct baryon targets:\,\,protons or electrons, allowing an exchange of energy and momentum. The resulting effect of such DM-baryon interactions is qualitatively similar in both cases:\,\,a suppression of $P(k)$ at scales depending on the interaction type and strength. 

To quantitatively incorporate the effects of DM-baryon interactions, collisional terms must be added to the cosmological perturbation equations to compute the evolution of linear perturbations in DM and baryon fluids. The formalism is well described in\,\cite{Dvorkin2014, Chen2002, Boddy2022, Gluscevic:2017ywp, Boddy2018, Boddy2018b, Li:2022mdj, Slatyer:2018aqg, Buen-Abad2022, He:2025npy, Boehm2005, Nguyen2021, Maamari2021, Nadler:2019zrb, DES:2020fxi, Xu:2018efh, Rogers2022, Becker2021} and detailed in Supplemental Material \cite{suppmat}.

We solve the modified cosmological perturbation equations for IDM scenarios using \texttt{class\_dmb}\,\cite{Buen-Abad2022}, a modified version of \texttt{CLASS}\,\cite{Blas2011}, to obtain $P(k)$. We recover both the suppression of power at relevant scales and acoustic features attributed to dark acoustic oscillations (DAO)\,\cite{Cyr-Racine:2013fsa, Verwohlt:2024efh, Zhang:2024mmg}.

\section{Model for the UV luminosity function}\label{sec:UVLF}

We describe the semi-analytical model for calculating the UVLF, a key observable of early galaxies, from $P(k)$. Two major steps are involved:\,\,{\it (i)} calculating the halo mass function (HMF) and {\it (ii)} prescribing the galaxy-halo connection. We further assume that each DM halo hosts one galaxy.

Since UVLF quantifies the number density of galaxies per unit UV magnitude, the relevant quantity is the HMF, defined as the number density of DM halos per unit mass at a given redshift $z$. Calculation of the HMF using the extended Press-Schechter formalism requires two functions calibrated from $N$-body simulations. We adopt the \textit{top-hat} (for ``JWST+" setup, smooth-$k$) filter function and use the ellipsoidal model of halo collapse, with the \textit{Sheth-Tormen} fitting function to compute the halo abundance. The effect of DM-baryon interactions is encoded in the mass variance $\sigma^{2}(M_h,z)$, where $M_{h}$ is the DM halo mass.

We use the galaxy-halo connection prescription from \textsc{Thesan-Zoom} simulations\,\cite{2025arXiv250300106M, 2025arXiv250220437K, Shen:2025isu, 2025arXiv250303806Z, 2025arXiv250302927Z, 2025arXiv250304894M, 2025arXiv250505554W, 2025arXiv250708787M, 2025arXiv251013977P}, a suite of zoom-in simulations tracing the evolution of galaxies during the first two billion years of the Universe, and thus suitable for JWST observations. We model the halo-scale star-formation efficiency (SFE), $\epsilon_{*}(M_h )$, by\,\cite{Shen:2025isu}
\begin{equation}
    \epsilon_\star(M_h)= \frac{\epsilon_{0}}{(M_h/M_{0})^{-\alpha_\star} + (M_h/M_{0})^{-\beta_\star}}\,,\label{eq:SFE}
\end{equation}
where $\epsilon_{0}$, $\alpha_{*}$, $\beta_{*}$, and $M_{0}$ are the SFE model parameters. This \textit{broken power law} model in\,\cite{Shen:2025isu} is attributed to distinct physical mechanisms that regulate gas outflows at such high redshifts.\,\,This model does not capture the turnover in the SFE at higher halo masses ($\sim 10^{12}M_{\odot}$) expected to be caused by AGN or other feedback processes\,\cite{2011IAUS..277..273S, 2022MNRAS.512.1052P}; instead, we cap $\epsilon_{\star}(M_{\mathrm{h}})$ at a constant value $\epsilon_{\star}(M_{c})$ beyond a critical halo mass $M_{c}$. This reflects the fact that the dynamical range probed by the \textsc{Thesan-Zoom} simulations extends only up to $M_h \leq M_c \simeq 10^{10}\,M_{\odot}$\,\cite{Shen:2025isu}. Throughout our analysis, we fix this capping mass at $\log_{10}{\left(M_{c}/M_{\odot}\right)} = 10.5$.\,\,We estimate the star formation rate (SFR) using the halo mass accretion model adopted in\,\cite{Shen:2025isu, 2024MNRAS.530.4868Y}, based on the \textsc{Gureft} simulations\,\cite{2024MNRAS.527.5929Y}.\,\,However, cosmologies with suppressed halo formation can modify the accretion history, making it sensitive to the underlying DM model. We discuss this point in the \textit{Supplemental Material} \cite{suppmat}.
Empirical relations between the SFR and specific UV luminosity of galaxies\,\cite{Shen2024, Shen:2025isu} allow us to calculate the median UV magnitude, $M_{\rm UV}$, as a function of $M_h$.

Some level of stochasticity is expected in various galaxy formation scenarios\,\cite{2017MNRAS.466...88S, 2016ApJ...833..254S, 2019ApJ...881...71E, 2020MNRAS.498..430I, 2020MNRAS.497..698T, 2021MNRAS.501.4812F, Shen2023}.
We model this effect by assuming a scatter of $\sim \mathcal{O}(1)~\rm{mag}$ (termed the UV \textit{variability}) about the median $M_{\mathrm{\rm UV}}-M_{h}$ relation.
We adopt the empirical model of variability in ref.\,\cite{Shen:2025isu} which matches the predictions from various simulations\,\cite{Feldmann:2024kwz, 2018MNRAS.479..994R, 2020MNRAS.494.2200K, 2018MNRAS.478.1694M, Sun2023a, Pallottini:2022inw, Pallottini2023}. Finally, we compute the binned UVLF $\bar{\Phi}^i_{\mathrm UV}$ by integrating the galaxy number counts over UV magnitude bins of width $\Delta M^{i}_{\mathrm{\rm UV}}$\,\cite{Sabti2024}
\begin{equation}
\bar{\Phi}^i_{\mathrm{\rm UV}}(z)
= \frac{1}{\Delta M^{i}_{\mathrm{\rm UV}}}
\int \mathrm{d}M'_{\mathrm{\rm UV}}
\int \mathrm{d}M_h \frac{\mathrm{d}n_h (z)}{\mathrm{d}M_h}\,
p(M'_{\mathrm{\rm UV}}| M_{h})\,,
\label{eq:UVLF_binned}
\end{equation}
where $n_{h}(z)$ is the comoving number density of halos, $dn_{h}(z)/dM_{h}$ is the HMF at redshift $z$, and $p(M_{\mathrm{\rm UV}}| M_{h})$ is the probability density that a halo of mass $M_h$ hosts a galaxy with UV magnitude $M_{\mathrm{\rm UV}}$. The outer integral is over the $i$-th $M_{\rm UV}$ bin. Following ref.\,\cite{Shen:2025isu}, we model $p(M_{\mathrm{\rm UV}}| M_{h})$ as a Gaussian with a mean $M_{\mathrm{\rm UV}}$ and width $\sigma_{\rm UV}(M_h)$, termed UV variability
\begin{eqnarray}
\sigma_{\rm UV}(M_h) = \text{max}[\sigma_{\rm UV, 10.5} - 0.34 \log_{10}(M_{h, 10.5}), 
0.4]\,, \phantom{xxx}
\end{eqnarray}
where $\sigma_{\rm{UV, 10.5}} $ is the UV variability at a DM halo mass of $10^{10.5}M_{\odot}$ and $M_{h, 10.5}$ is the halo mass scaled by $10^{10.5}M_{\odot}$. Our astrophysical model is assumed to be independent of $z$, so the entire $z$-dependence in UVLF arises from the HMF.

In the next section, we discuss how we use this model to constrain the DM-baryon scattering cross-section.

\section{Results}
\label{sec:results}

We sample the model parameters using the Markov Chain Monte Carlo (MCMC) code \textsc{MontePython}\,\cite{2019PDU....24..260B, 2018ascl.soft05027B, Audren:2012wb} 
with the \textsc{Gallumi}\,\cite{2022PhRvD.105d3518S} (modified according to\,\cite{Shen:2025isu}) and \textsc{Pantheon} likelihood pipelines. We discuss our choice of priors in the \textit{End Matter}.

The constraining power of a particular observable on IDM is determined by some key factors, such as scales probed by the observable, the statistical and systematic uncertainties in the data, as well as possible discrepancies between the $\Lambda$CDM model and the data. For example, it is possible for suppression effects caused by IDM to be compensated by unaccounted-for systematics in the astrophysical model, making it difficult to isolate the impact of the interaction.

The highest-redshift galaxy spectroscopically confirmed by JWST lies at  $z = 14.44 \pm 0.02$\,\cite{2025arXiv250511263N}. In fig.~\ref{fig:uvlf-transfer-proton} (right panel), we present UVLFs computed for $\Lambda$CDM and IDM cosmologies at $z \approx 14$. As expected, the UVLFs calculated for IDM cosmologies are suppressed with respect to those of $\Lambda$CDM. 
The UVLF for the $\Lambda$CDM cosmology, with astrophysical parameters ($\epsilon_{0}$, $\alpha_{*}$, $\beta_{*}$, $M_{0}$, and $\sigma_{\mathrm{UV}}(10^{10.5} M_{\odot})$) set to the best-fit values given in\,\cite{Shen:2025isu}, fits the data better compared to the IDM scenarios.

\begin{figure}[!h]
    \centering
    \includegraphics[width=1.0\columnwidth]{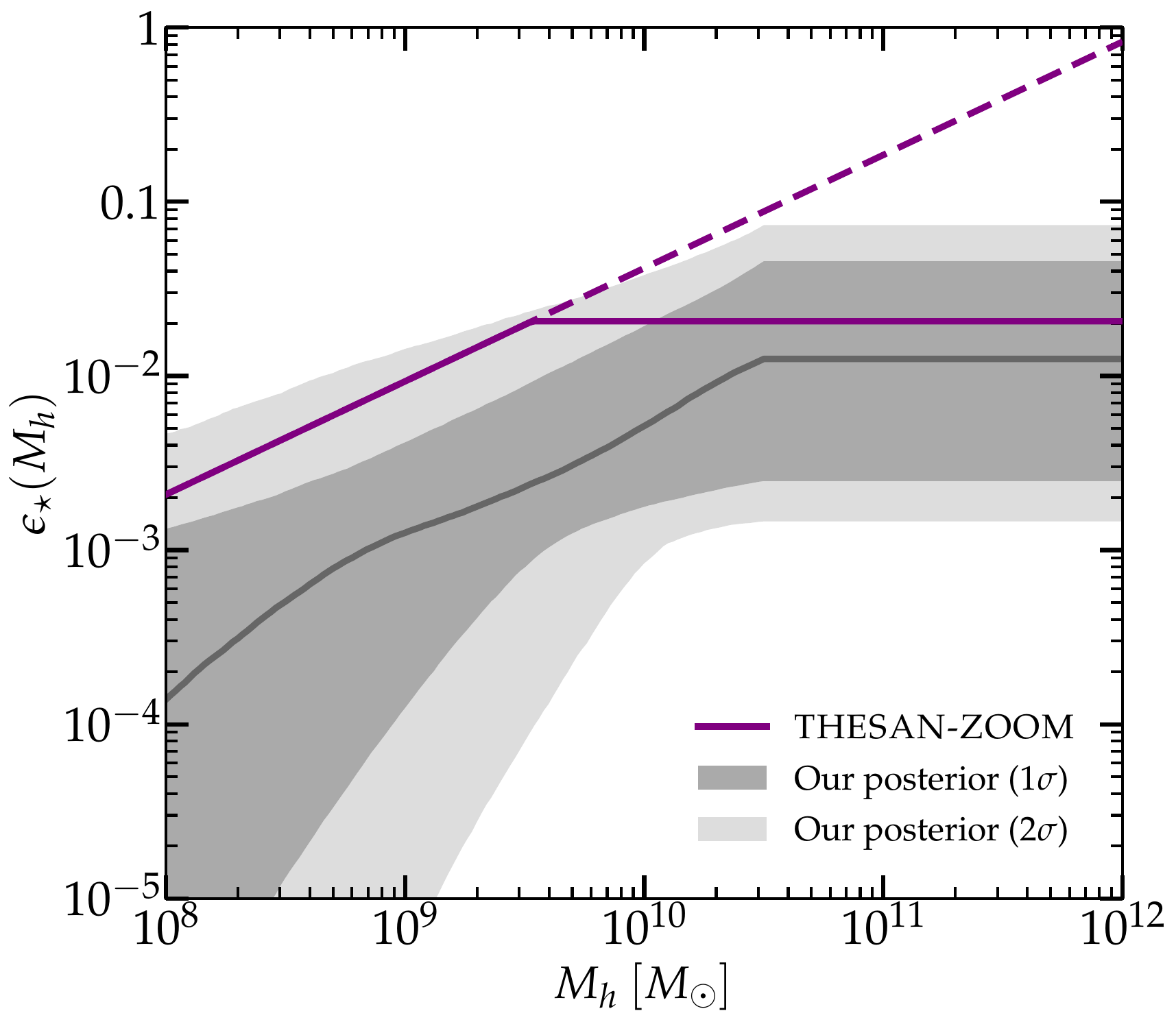}
    \caption{
The \textbf{magenta} solid line shows the SFE as a function of halo mass for the $\Lambda$CDM cosmology, using the best-fit SFE model parameters from\,\cite{Shen:2025isu} at $z\approx 14$. 
The \textbf{gray line}, and the \textbf{dark gray} and \textbf{light gray} shaded regions show respectively the median, 1$\sigma$, and 2$\sigma$ intervals for SFE (Eq.\,\ref{eq:SFE}) sampled from our posterior pdfs for $m_\chi=1~\rm{MeV}$, $n=-2$, $B=p$, and with ``JWST Spec-z" setup as a function of $M_h$. The dashed line represents the $\epsilon_\star(M_{h})$ from the \textsc{Thesan-Zoom} best-fit model, extended beyond the cutoff $M_{c} = 10^{10.5}\,M_{\odot}$}
    \label{fig:SFE-posterior}
\end{figure}

The ordering of UVLFs for different values of $n$ in fig.~\ref{fig:uvlf-transfer-proton} (right panel)  recovers the corresponding order of the $P(k)$ (left panel) at $k \approx 10\,\mathrm{Mpc}^{-1}$ (shown by the gray shaded region in the left panel) corresponding to the smallest scales probed by the JWST UVLF\,\cite{2024ApJ...976...40W}.
Because the suppression due to IDM shifts to larger $k$ for smaller $\sigma^{\chi B}_{n}$, the maximum $k$ values probed by an observation are important in determining the stringency of the corresponding upper limits. The one-to-one correspondence between transfer functions and UVLFs across different $n$ values may not hold, as each case involves distinct astrophysical parameters. We show similar plots for DM-electron interaction in the \textit{Supplemental Material} \cite{suppmat}.

In fig.~\ref{fig:SFE-posterior}, we present the best fit, as well as the 1$\sigma$ and 2$\sigma$ bands on SFE (in gray bands) as a function of halo mass $M_{h}$, derived from MCMC samples of the model parameters, for DM-proton interaction, $n=-2$ case, with $m_{\chi} = 1\,$MeV with ``JWST Spec-z" setup. The magenta line, representing the best-fit SFE model parameters from \cite{Shen:2025isu}, remains consistent within the 2$\sigma$ confidence interval, and at higher halo masses, within the 1$\sigma$ interval. This concordance suggests that the \textsc{Thesan-Zoom} simulation-based model reliably captures the star formation physics relevant for JWST observations.

We show our limits for the mass range $1\,\mathrm{MeV}-1\,\mathrm{GeV}$.  Cosmological probes like the Lyman-$\alpha$ forest measurements\,\cite{Cen:1994da}, CMB lensing\,\cite{Amblard:2004ih}, MWS \cite{Nadler:2019zrb}, and JWST (this work) give strong constraints on sub-GeV DM masses. In contrast, direct detection experiments are more sensitive to heavier DM masses\,\cite{Orrigo:2015cha, XENON:2017vdw, XENON:2023cxc, LUX:2013afz, Szydagis:2016few, Zhang:2025ajc, CDMS-II:2004xnm, Rau:2006nw, Caminata:2019xwi, Scorza:2009nhq, Agnes:2023izm, COSINE-100:2023tcq, Petriello:2008jj,Cao:2019rap}. 

We do not consider DM particle masses below 1 MeV since Big Bang nucleosynthesis (BBN) and CMB place the strongest constraints for this mass range by measuring the  cosmological relativistic degrees of freedom\,\cite{Sabti:2019mhn, Creque-Sarbinowski:2019mcm, Munoz:2018pzp}. In IDM cosmologies, during very early epochs, DM interacts with baryons, thermally populating the baryonic plasma\,\cite{Dicus:1982bz}. For strong $\sigma^{\chi B}_{n}$ relevant to this work, this can increase the number of relativistic degrees of freedom, $N_{\mathrm{eff}}$, during BBN and CMB epochs, which is disallowed, given that there are three neutrino species.

\section{Discussions and Conclusions}
We investigate DM–baryon interactions using JWST high-redshift UVLF data. For a broad class of models, the momentum-transfer interaction cross-section between DM and baryons can be expressed as $\sigma^{\chi B}_{T,n} = \sigma^{\chi\text{}B}_{n} v^{n}$. We constrain $\sigma^{\chi B}_{n}$ for $n = 0, -2,$ and $-4$, with $B \in \{p, e\}$.
From fig.~\ref{fig:n_minus_2_proton_Spec_z}, we see that JWST provides the strongest constraint across a wide range of sub-GeV DM masses for the $ n = -2$ case. This is attributed to the nature of suppression for this case, where the transfer function diminishes gradually, making the suppression at JWST scales comparable to those at Lyman$-\alpha$ and MWS scales. This is a new observable in addition to existing probes of DM-baryon scattering\,\cite{Cirelli2024, Bell:2023sdq, LZ:2025iaw, Bhoonah:2018gjb, An:2017ojc, Wadekar:2019mpc, Bell:2019egg, Granelli:2022ysi, Cappiello:2019qsw, Dent:2020qev, DAmico:2009tep, Boehm2001, Boehm2005, PhysRevLett.126.091101, Maamari2021, Boddy2018, Becker2021,  Nadler:2019zrb,  Zhou:2024igb, Vogelsberger:2015gpr, Esteban2023, Crumrine2024, He:2025npy, Gluscevic:2017ywp, Gluscevic:2019yal,Hooper:2022byl, An:2021qdl, Dhyani:2025uof, Zhang:2024mmg, Herrera:2023nww, Bhutani:2025jfo, Maity:2022exk}.

The largest comoving wavenumbers $k$ probed by cosmological observations follow $ k_{\mathrm{CMB}} \lesssim k_{\mathrm{Ly\text{-}\alpha}} \lesssim k_{\mathrm{JWST}}  \lesssim k_{\mathrm{MWS}} $, corresponding to the CMB, Lyman-$\alpha$ forest, JWST, and MWS observations, respectively. However, the constraining power of a cosmological observation depends on both the dataset and the nature of the suppression at the largest $k$ scales it probes. Due to these factors, for $n = 0$, the resulting limits are weaker than those from the Lyman-$\alpha$ and MWS observations. We show the corresponding constraint plots in the \textit{Supplemental Material}\cite{suppmat}. 

Even though we used the smooth-$k$ filter function in our ``JWST+'' setup, we find that applying this filter function only to the spectroscopic data used in the ``JWST Spec-z" setup does not lead to any substantial change in our results compared to using the \textit{top-hat} filter function. This is possibly attributed to the sparse nature of the data, which is expected to change as we observe more galaxies at these redshifts. However, the suppression of smaller structures can change the merger rate in the early Universe, as in such cases the number of smaller halos decreases. This can lead to reduced halo accretion rates which become relevant for warm dark matter (WDM) or IDM scenarios. We study the effect of suppression in accretion rate for $n = -2$ case for the ``JWST+" setup and present our results in the \textit{Supplemental Material} \cite{suppmat}.

The UVLF data carry information not only about the cosmological model, but also about the astrophysics at such high redshifts. Pre-JWST astrophysical models within the $\Lambda$CDM cosmology have failed to explain the JWST data on galaxy abundance\,\cite{Parashari2023, Sabti2024,  Boylan-Kolchin:2022kae}. To ensure the robustness of our results, we marginalize over the astrophysical parameters so that the limits become insensitive to the choice of astrophysical modeling.

Our work presents, {\it for the first time}, new regions of DM-baryon scattering cross-sections probed using JWST observations. Although the currently available dataset provides the strongest constraints only on dipole-moment-like ($n = -2$) DM-proton interactions, the discovery of more galaxies at high redshifts is expected to make the limits for other values of $n$ competitive with existing bounds, and possibly even the most stringent. In addition, understanding the astrophysics at the cosmic dawn through simulations adapted for high redshifts will lead to insightful conclusions, including the possibility of new discoveries. Through observations of the first galaxies, we hope to explore novel ways to probe the fundamental nature of DM.

\vskip 2 em

\textbf{Note added:} Recently, Ref.\,\cite{Lazare:2025gha} appeared on arXiv, exploring a similar scenario using the HST UVLF dataset.\,\,We use the JWST UVLF measurements, which contain more high-redshift data compared to HST. At higher redshifts, structures become smaller and probe larger values of $k$ and, thus, become sensitive to lower values of the DM - baryon scattering cross-sections. JWST is, therefore, capable of providing the strongest constraint on one class of DM--proton interaction models using only $\sim 40$ galaxies. As new data arrives, it is expected to place the best constraints on a broader class of models. Their paper studied the DM-proton interactions only, while we present constraints on both DM-proton and DM-electron interactions. While they studied $n = 0, +2, +4$ models, we study $n = 0, -2, -4$ models.

\label{sec:Conclusion}
\begin{acknowledgments}

\section{Acknowledgments}

We especially thank Priyank Parashari for initial discussions that led to this project. We especially thank Anirban Das, Vera Gluscevic, and Priyank Parashari for detailed discussions about our work. We thank Susmita Adhikari, Arka Banerjee, Debajit Bose, Subhadip Bouri, Ariane Dekker, Deep Jyoti Das, Prakash Gaikwad, Isabel Garcia Garcia, Durba Ghosh, Anindya Guria, Mark Krumholz, Nirmal Raj, Tirthankar Roy Chowdhury, Akash Kumar Saha, Martin Schmaltz, Yashi Tiwari, Ujjwal Kumar Upadhyay, Ken Van Tilburg, and Harrison Winch for discussions and useful comments. We are grateful to Gaurav Narain for kindly allowing us to use his workstations for performing the \textsc{MontePython} MCMC runs.
S.D.\,\,acknowledges financial support from Kishore Vaigyanik Protsahan Yojana (KVPY) and from the Ohio State University via University Fellowship.
R. M.\,\, is supported by the ISRO-IISc STC Grant No. ISTC/PHY/RL/499 and ANRF National Postdoctoral Fellowship (NPDF) under Project No. PDF/2025/001161.
A. S.\,\, acknowledges the Ministry of Human Resource Development, Government of India, for financial support via the Prime Ministers’ Research Fellowship (PMRF). 
R. L.\,\, acknowledges financial support from the institute start-up funds, ISRO-IISc STC for Grant No. ISTC/PHY/RL/499, and ANRF for Grant No. ANRF/ARG/2025/005140/PS. We would like to thank Vinayak Sanil, who maintained the computing workstations, which was invaluable for various computations performed in this work. R. L. would like to acknowledge support from the ICTP through the Associates Programme and from the Simons Foundation, whose grant (Record ID: SFI-MPS-T-Institutes-00012057, AD) also supported this work.
\end{acknowledgments}

\newpage

\bibliographystyle{JHEP}
\bibliography{refs.bib}

\clearpage

\section{End Matter}

\renewcommand{\theequation}{A\arabic{equation}} 
\setcounter{equation}{0} 
\renewcommand{\thetable}{A\arabic{table}} 
\setcounter{table}{0} 

\subsection{UVLF Data}

The UVLF data used in this paper are taken from a set of 25 galaxies presented in \cite{Harikane2024a}, 10 galaxies from ref.\,\cite{Harikane2024b} (and references therein), 10 galaxies from\,\cite{Fujimoto:2023fgh}, 7 galaxies from\,\cite{2025A&A...693A..50N}, 2 galaxies from ref.\,\cite{2025arXiv250511263N}, 2 galaxies from\,\cite{2025ApJ...983L..22K}, 5 galaxies from\,\cite{2025arXiv250405893C}, 9 sources from\,\cite{Perez-Gonzalez:2025bqr}, and upper limits on UVLFs from\,\cite{2025arXiv250405893C, 2025arXiv250100984W, 2025arXiv250706292W}. We only consider UVLF measurements at redshifts $z \gtrsim 8$. All data below redshift $z \sim 16$ are obtained from spectroscopically confirmed galaxies, while for our ``JWST+" analyses, we also include photometric ultra-high-redshift UVLF data and upper/\,lower limits for $z \gtrsim 17$. At each redshift, we convert the set of discrete $M_{\rm UV}$ values to bins either using the bin widths quoted in the respective paper or conservatively by choosing the larger of the differences to adjacent $M_{\rm UV}$ values as the bin width.

\subsection{Likelihood Analysis}

We modified part of the \textsc{Gallumi} pipeline, adopting the formalism of ref.\,\cite{Shen:2025isu} as described in the text. 

We incorporate dust using the prescription of ref.\,\cite{2016ApJ...833...72B}, extrapolated to $z \approx 10$, which matches closely the \textsc{Thesan-Zoom} simulations\,\cite{2025arXiv250220437K}.
We use the following likelihood function for the binned-$\bar{\Phi}_{\rm UV}$ in each bin:

\begin{equation}
\log \mathcal{L}
=-\frac{1}{2}\sum_{z}\sum_{i}
\begin{cases}
\qty(\dfrac{\bar{\Phi}^i_{\rm UV}-\Phi^i_{\rm UV}}
{\delta\Phi_{\rm UV}^{{\rm up}, i}})^{2},
& \bar{\Phi}^i_{\rm UV}>\Phi^i_{\rm UV}\\
\qty(\dfrac{\bar{\Phi}^i_{\rm UV}-\Phi^i_{\rm UV}}
{\delta\Phi_{\rm UV}^{{\rm lo}, i}})^{2},
&  \bar{\Phi}^i_{\rm UV}<\Phi^i_{\rm UV}
\end{cases}
\end{equation}
where $\bar{\Phi}^i_{\rm UV}$ is the binned UVLF calculated in eq.~\ref{eq:UVLF_binned}, and is thus a function of the cosmological model and astrophysical parameters. Moreover, it implicitly depends on $z$ through its dependence on the HMF. The JWST measurement (with $1\sigma$ uncertainties) of the UVLF in the $i$-th bin at redshift $z$ is given by 
$$\qty(\Phi^i_{\rm UV})^{+ \delta\Phi_{\rm UV}^{{\rm up}, i}}_{-\delta\Phi_{\rm UV}^{{\rm lo}, i}}$$
Our likelihood function enables accommodation of asymmetric error bars on the measured UVLFs. We sum over each $M_{\rm UV}$ bin at all redshifts, relevant to different analysis.

We run the likelihood analysis for different choices of DM mass $m_\chi$, baryon component $B$ and velocity-dependence\,$n$.\,\,The free parameters of our model include cosmological parameters $\{n_s, A_s, \omega_b, \Omega_\chi, \sigma_n^{\chi B}\}$, and astrophysical (nuisance) parameters $\{\alpha_\star, \beta_\star, \epsilon_0, M_0, \sigma_{\rm UV}(10^{10.5}M_\odot), M\}$. Here $n_s$, $A_s$ and $\omega_b$ are standard $\Lambda$CDM parameters while $\Omega_\chi$ replaces the CDM abundance $\Omega_c$ (we assume 100\% IDM). We impose Gaussian priors on $\omega_b$ from BBN\,\cite{2009PhR...472....1I}, and on $A_{s}$, $n_{s}$ from CMB measurements\,\cite{Planck:2018vyg}, since JWST UVLF data cannot constrain them stringently\,\cite{2024ApJ...976...40W}.\,\,Further, we fix $\theta_s$ and $\tau$ to their \textit{Planck} measurements\,\cite{Planck:2018vyg}.
We assume uniform priors on the DM abundance $\Omega_\chi$, logarithm of the cross-section $\log_{10}\sigma_n^{\chi B}$, parameters $\{\alpha_\star, \beta_\star, \log_{10}\epsilon_0, \log_{10}M_0\}$ characterizing the SFE, the UV variability $\sigma_{\rm UV}(10^{10.5}M_\odot)$, and the intrinsic SNIa magnitude $M$. These priors are summarized in Table \ref{tab:priors_astro_cosmo}. 
Note that the prior on $\log_{10}\sigma_n^{\chi-B}$ is always uniform, but the range is chosen according to the specific choice of $(m_\chi, B, n)$.

\begin{table}[ht]
\caption{Priors on cosmological and astrophysical parameters. Ranges for the astrophysical parameters are motivated by the \textsc{Thesan-Zoom} simulations\,\cite{Shen:2025isu} }
\label{tab:priors_astro_cosmo}
\begin{ruledtabular}
\begin{tabular}{lc}
Parameter & Prior \\
\hline \\
$\log_{10}\sigma_n^{\chi B}$ & Uniform\footnote{The range of the uniform prior is selected according to the choice of $(m_\chi, B, n)$.} \\
$\Omega_{\chi}$            & $\mathcal{U}[0.0,\,0.3]$               \\
$\omega_b \times 10^{2}$      & $\mathcal{N}(2.233,\,0.036)$ \\
$n_{s}$      & $\mathcal{N}(0.9649,\,0.0042)$ \\
$A_{s}\times10^{9}$      & $\mathcal{N}(2.1,\,0.03)$ \\~\\
\hline \\
$\alpha_\star$                & $\mathcal{U}[0.0,\,3.0]$              \\
$\beta_\star$                 & $\mathcal{U}[0.0,\,3.0]$               \\
$\log_{10}{\epsilon_{0}}$    &
$\mathcal{U}[-3.0,\,0.0]$              \\
$\log_{10}M_{0}$             &
$\mathcal{U}[7.0,\,11.0]$             \\
$\sigma_{\rm UV}(10^{10.5} M_{\odot})$ & 
$\mathcal{U}[0.001,\,3.0]$ \\
$M$                           & $\mathcal{U}[-21,\,-18]$        \\~\\
\end{tabular}
\end{ruledtabular}
\end{table}

As highlighted in the main text, we used \textsc{MontePython} to perform Markov Chain Monte Carlo (MCMC) runs with the likelihood and prior as discussed above. We run 40 parallel MCMC chains until the Gelman-Rubin criterion for each parameter satisfies $R<1.05$. We marginalize the posterior sample to obtain 95\% C.L. upper limits on $\log_{10}(\sigma_n^{\chi B})$. This procedure is repeated for each individual choice of $m_\chi$, $B$ and $n$. We vary over $n\in\{0, -2, -4\}$, $B \in \{e, p\}$ and 5 choices of $m_\chi$ in the range $1~{\rm MeV} \leq m_\chi \leq 1~{\rm GeV}$.

\newpage

\onecolumngrid
\begin{center}
\textbf{\large \mytitle}

\vspace{0.08in}
{ \it \large Supplemental Material}\\ 
\vspace{0.12in}

{Souradeep Das, Ranjini Mondol, Abhijeet Singh, and Ranjan Laha}
\end{center}
\twocolumngrid
 \setcounter{equation}{0}
 \setcounter{figure}{0}
\setcounter{section}{0}
 \makeatletter
 \renewcommand{\theequation}{S\arabic{equation}}
 \renewcommand{\thefigure}{S\arabic{figure}}

\section{Constraint on DM-baryon scattering from JWST UVLF data}

In this work, we obtained the upper limits on the normalization, $\sigma^{\chi B}_{n}$, of dark matter (DM)-baryon scattering for the cases $n = 0$, $n = -2$, and $n = -4$ with $B \in \{p, e\}$. We show the 95\% confidence level (C.L.) on $\sigma^{\chi p}_{-2}$ in the main text. The rest of the constraints are presented in this \textit{Supplemental Material} in Figs. \ref{fig:constraint-proton-electron-n-0} - \ref{fig:constraint-proton-electron-n-4}. 

\begin{figure*}
	\begin{center}
		\includegraphics[width=0.9475\columnwidth]{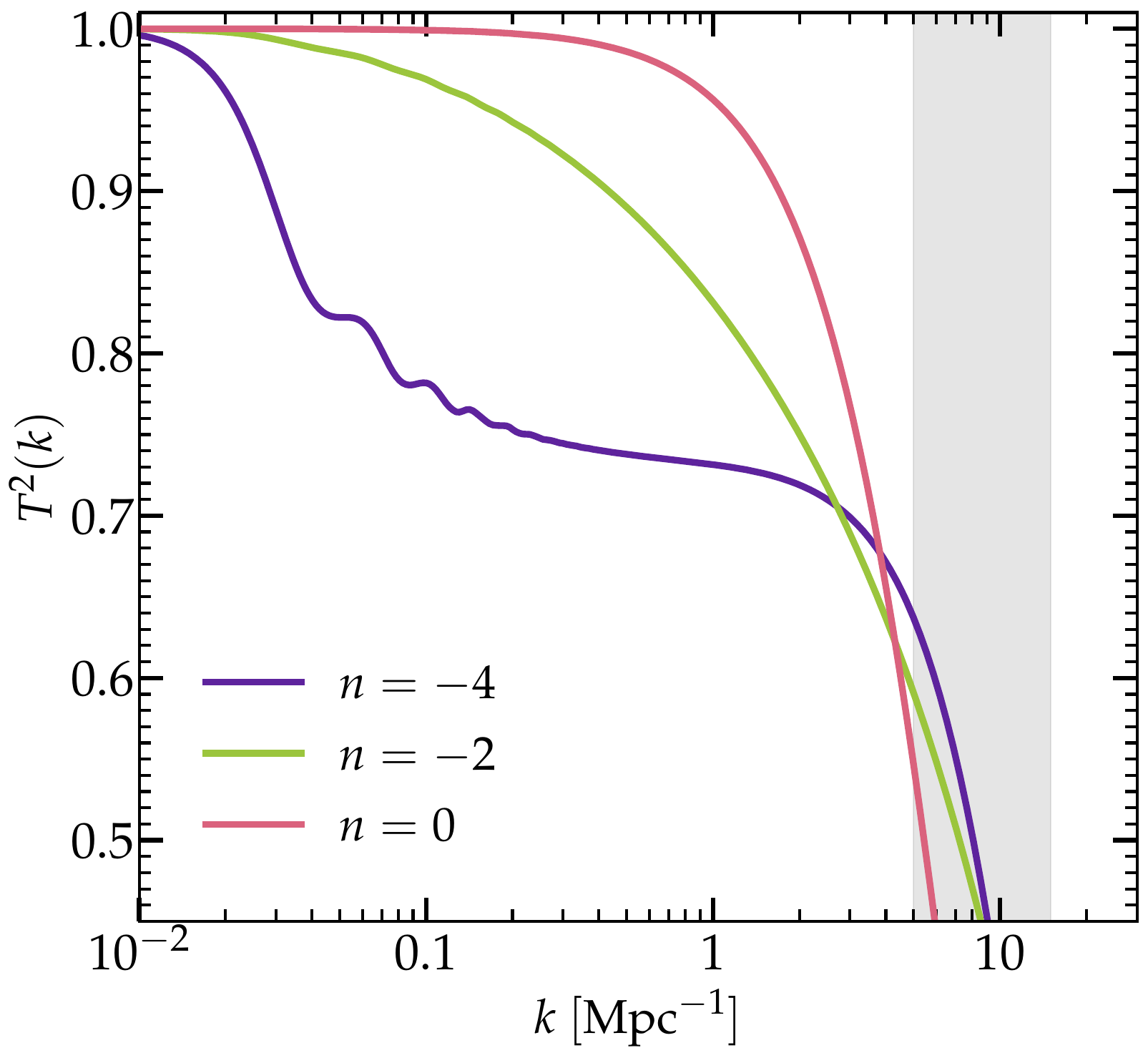}~~
		\includegraphics[width=\columnwidth]{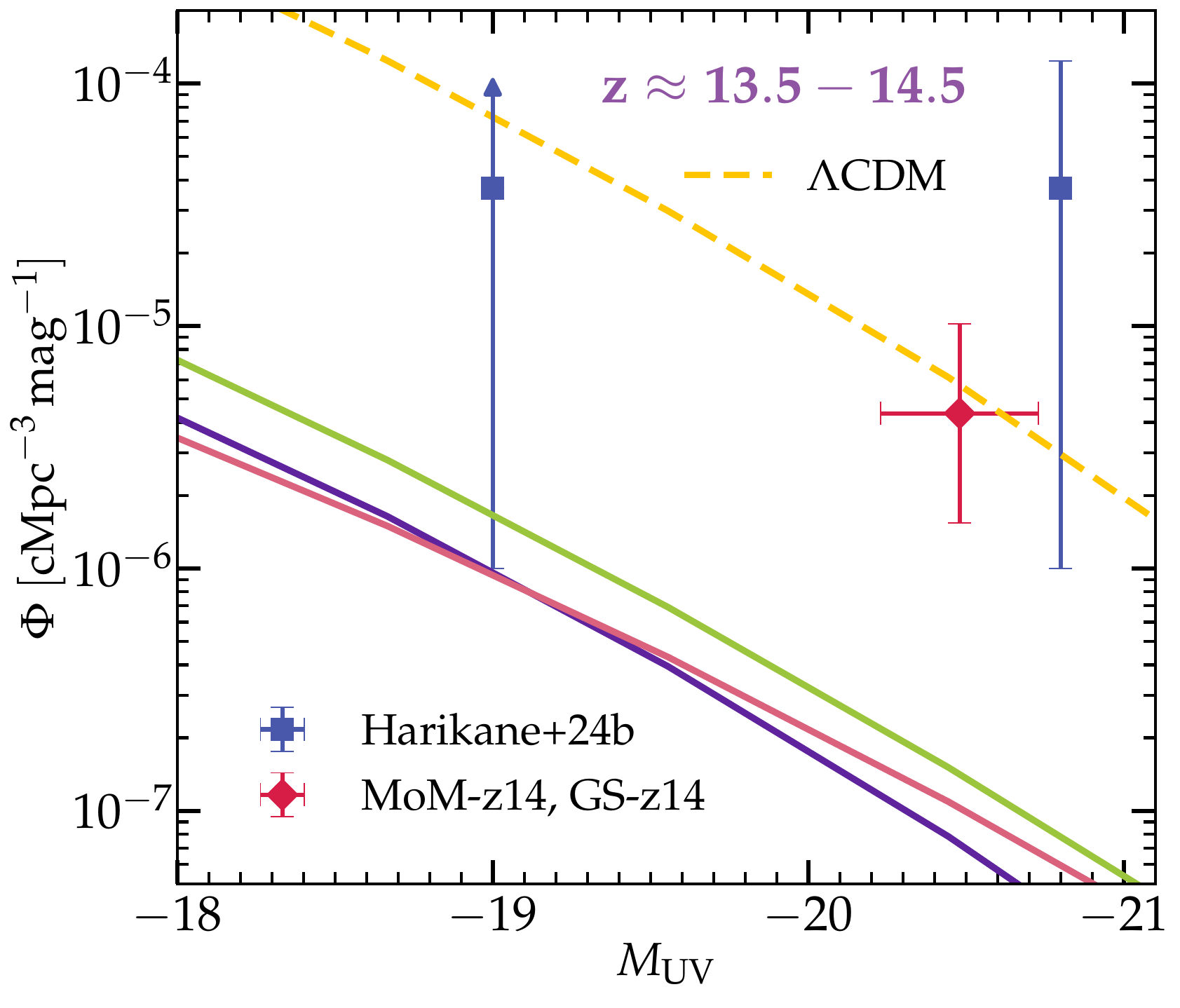}~~\\	
		\caption{
\textbf{\textit{Left:}} Comparison of the ratio $T^2(k) = P_{\rm IDM}(k)/P_{\Lambda\rm CDM}(k)$ of power spectrum of IDM to that of non-interacting $\Lambda$CDM. We consider IDM cases with different values of $n$: $n = 0$ (\textbf{pink}), $n = -2$ (\textbf{green}), and $n = -4$ (\textbf{purple}), for DM of mass $m_\chi = 1~\rm{MeV}$, interacting with electrons. 
Cross-section for each case corresponds to the 95\% C.L. upper limits obtained in this work: $\log_{10}(\sigma^{\chi e}_{n}/\mathrm{cm}^{2})$ for $n = 0,-2,-4$ cases are  $-27.84$, $-31.14$, and $-35.42$, respectively using ``JWST Spec-z" analysis. Suppression of power spectrum at the characteristic scales (\textbf{gray shaded region}) is visible.
\textbf{\textit{Right:}} Comparison of UVLF at $z\approx 14$ for $\Lambda$CDM (\textbf{dashed yellow}) and IDM (same color code as \textit{left} panel) for the same IDM scenarios. Data points represent UVLF measurements at $z \approx 14$ as derived in Harikane et al.\,\cite{Harikane2024b} (\textbf{blue squares}) and for the measurement of the MoM-z14 + GS-z14 galaxy observations\,\cite{2025arXiv250511263N}(\textbf{red diamond}). The choice of astrophysical parameters for $\Lambda$CDM correspond to the best-fit values from\,\cite{Shen:2025isu} and $\log_{10}{(M_{c}/M_\odot)} = 10.5$, while those for the interacting scenarios are fixed at the best-fit values obtained from our model by MCMC sampling. 
    }
		\label{fig:uvlf-transfer-electron}
	\end{center}	
\end{figure*}

\begin{figure*}[htb]
	\begin{center}
		\includegraphics[width=\columnwidth]{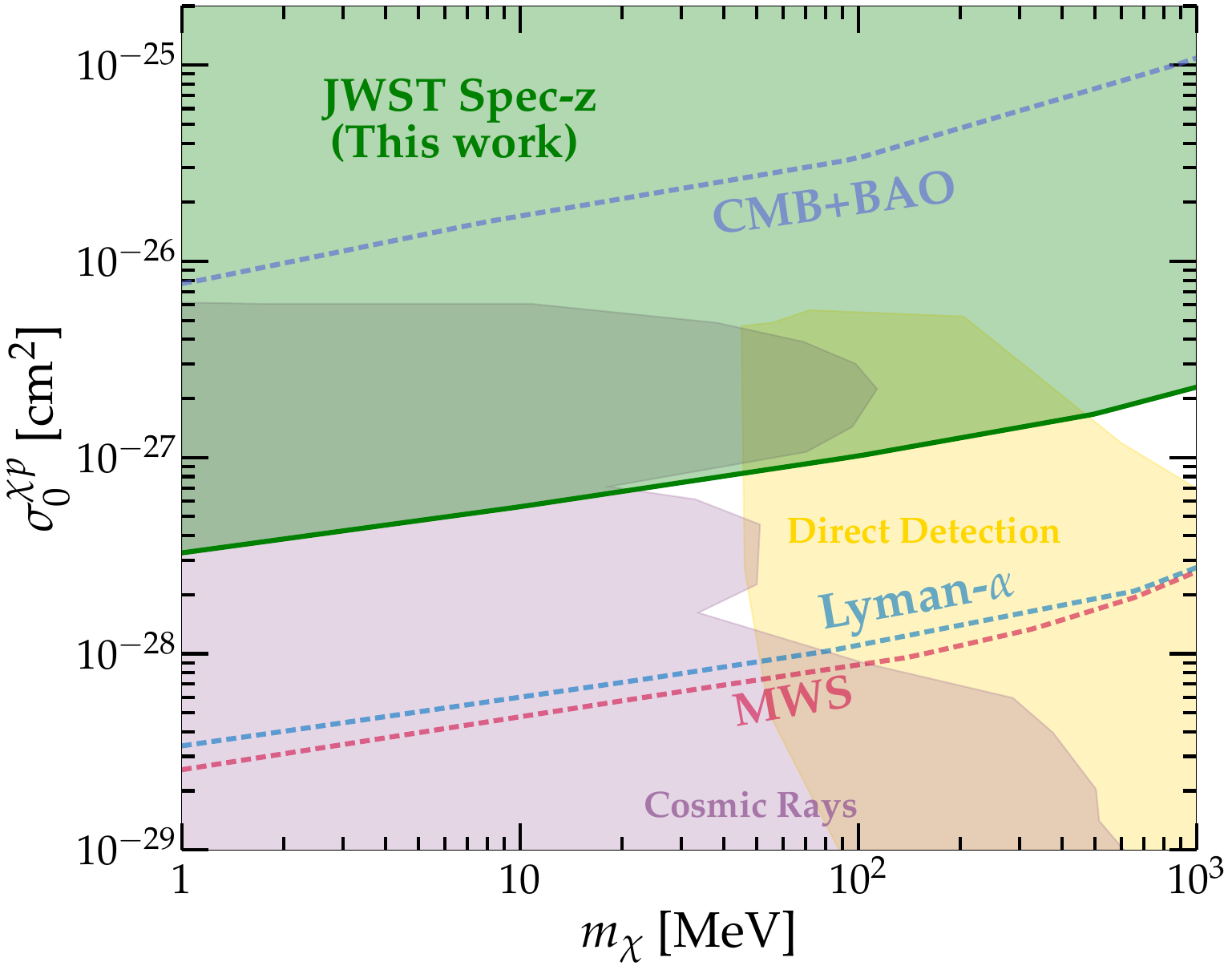}~~
		\includegraphics[width=\columnwidth]{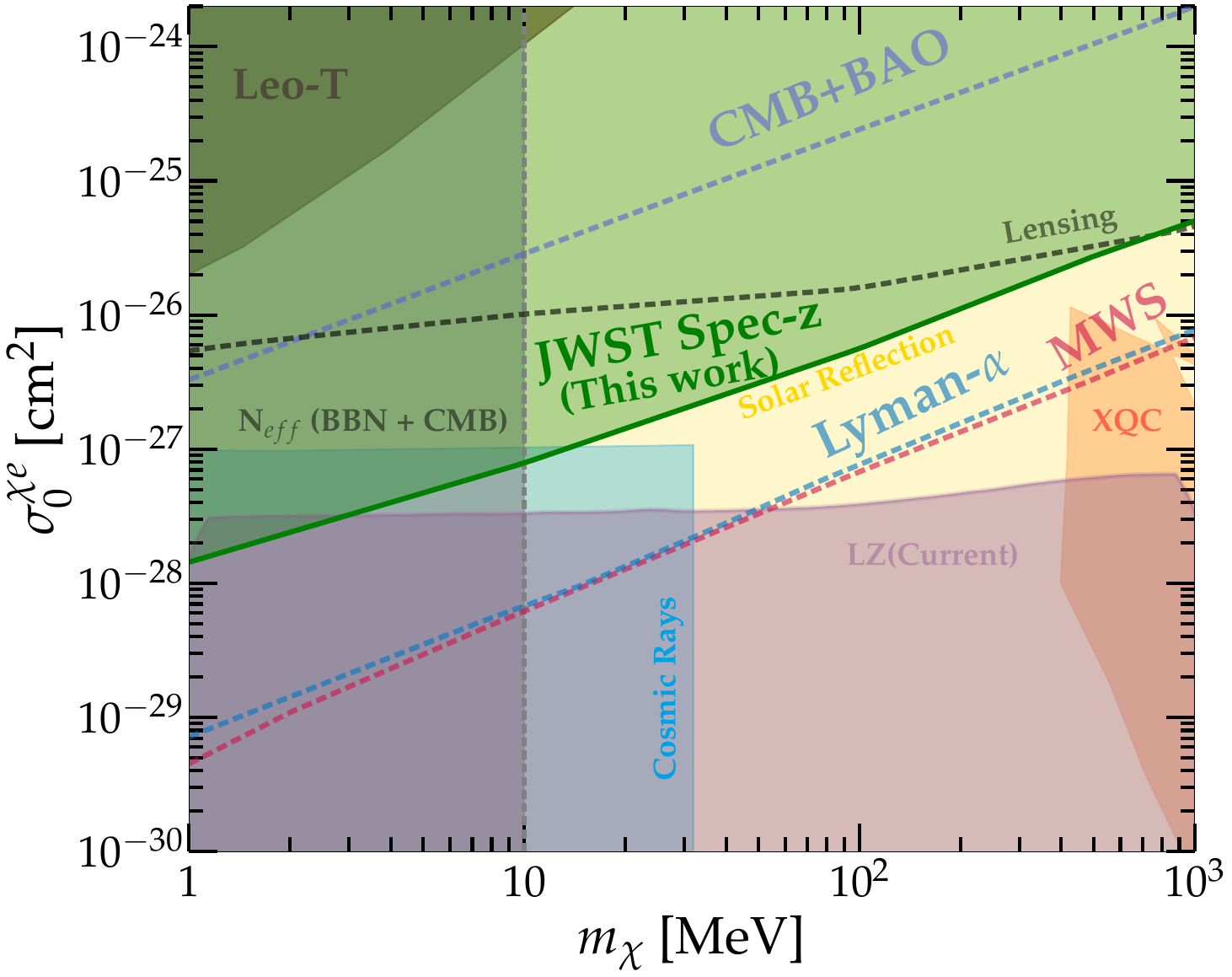}~~\\	
		\caption{95\% C.L.\,\,upper limits on the velocity-independent DM-baryon elastic scattering cross-section as a function of the DM mass $m_\chi$. The \textbf{green} shaded region in both panels is ruled out by the ``JWST Spec-z" setup (our work). The various dashed curves represent other cosmological constraints -- Lyman-$\alpha$ forest (cyan), Milky Way satellite count (pink), and CMB+BAO (purple)\,\cite{Buen-Abad2022, Nguyen2021}.
        \textbf{\textit{Left:}} Shows constraints on DM-proton interactions. Dark pink shaded region represents the parameter space excluded by cosmic ray observations \cite{Yin:2018yjn}. Yellow shaded area shows the region ruled out by direction detection experiments\,\cite{CRESST:2019jnq, CRESST:2019axx, CRESST:2017ues, XENON:2017vdw, EDELWEISS:2019vjv}. Some other constraints on $\sigma^{\chi p}_{0}$ are shown in\,\cite{Buen-Abad2022}.
        \textbf{\textit{Right:}} Shows constraints on DM-electron interactions. The celestial blue shaded region represents constraints imposed by cosmic ray observations\,\cite{Ema:2018bih, Cappiello:2019qsw}. Yellow shaded region represents constraints from solar reflection of cosmic rays\,\cite{An:2017ojc, Emken:2021lgc}. Brown shaded region shows constraints obtained from gas cooling in Leo-T dwarf galaxy\,\cite{Wadekar:2019mpc}. The shaded gray region shows constraints from BBN + CMB\,\cite{Sabti:2019mhn}. The orange shaded region is constrained from XQC rocket\,\cite{Erickcek:2007jv, Mahdawi:2018euy}. The black dashed line shows the constraints from weak lensing\,\cite{Zhang:2024mmg}. The dark pink  shaded region is constrained from LUX-ZEPLIN (LZ) collaboration\,\cite{Maity:2022exk}. Some other constraints on $\sigma^{\chi e}_{0}$ are shown in\,\cite{Buen-Abad2022,An:2021qdl}. 
    }
		\label{fig:constraint-proton-electron-n-0}
	\end{center}	
\end{figure*}

\begin{figure}[htb]
    \centering
    \includegraphics[width=1.0\columnwidth]{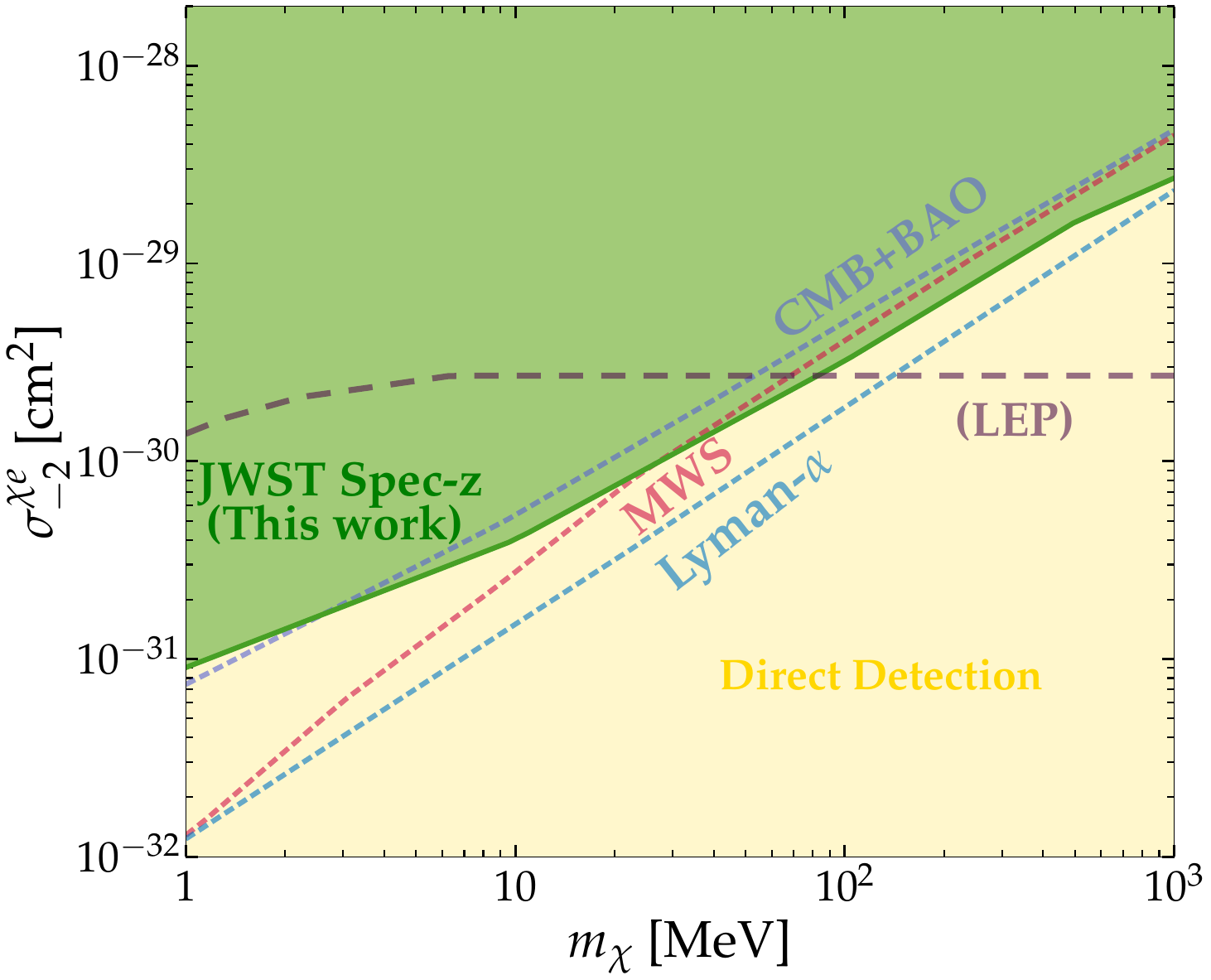}
    \caption{95\% C.L.\,\,upper limits on the normalization of velocity-dependent ($\sigma\propto v^{-2}$) DM-electron elastic scattering cross-section as a function of the DM mass $m_\chi$. The \textbf{green} shaded region is ruled out by the``JWST Spec-z" setup (our work). The various dashed curves represent other cosmological constraints -- Lyman-$\alpha$ forest (cyan), Milky Way satellite count (pink), and CMB+BAO (purple)\,\cite{Buen-Abad2022, Nguyen2021}. The yellow shaded region shows the regions excluded by Direct Detection experiments\,\cite{Essig:2012yx, Essig:2017kqs, XENON10:2011prx, XENON:2016jmt, DarkSide:2018ppu, SuperCDMS:2018mne, Crisler:2018gci, SENSEI:2019ibb, SENSEI:2020dpa}. The brown dashed line corresponds to the upper limits from LEP \cite{Fortin:2011hv}. Some other constraints on $\sigma^{\chi e}_{-2}$ are shown in\,\cite{Buen-Abad2022}.
    }
    \label{fig:constraint-electron-n-2}
\end{figure}

\begin{figure*}[htb]
	\begin{center}
		\includegraphics[width=\columnwidth]{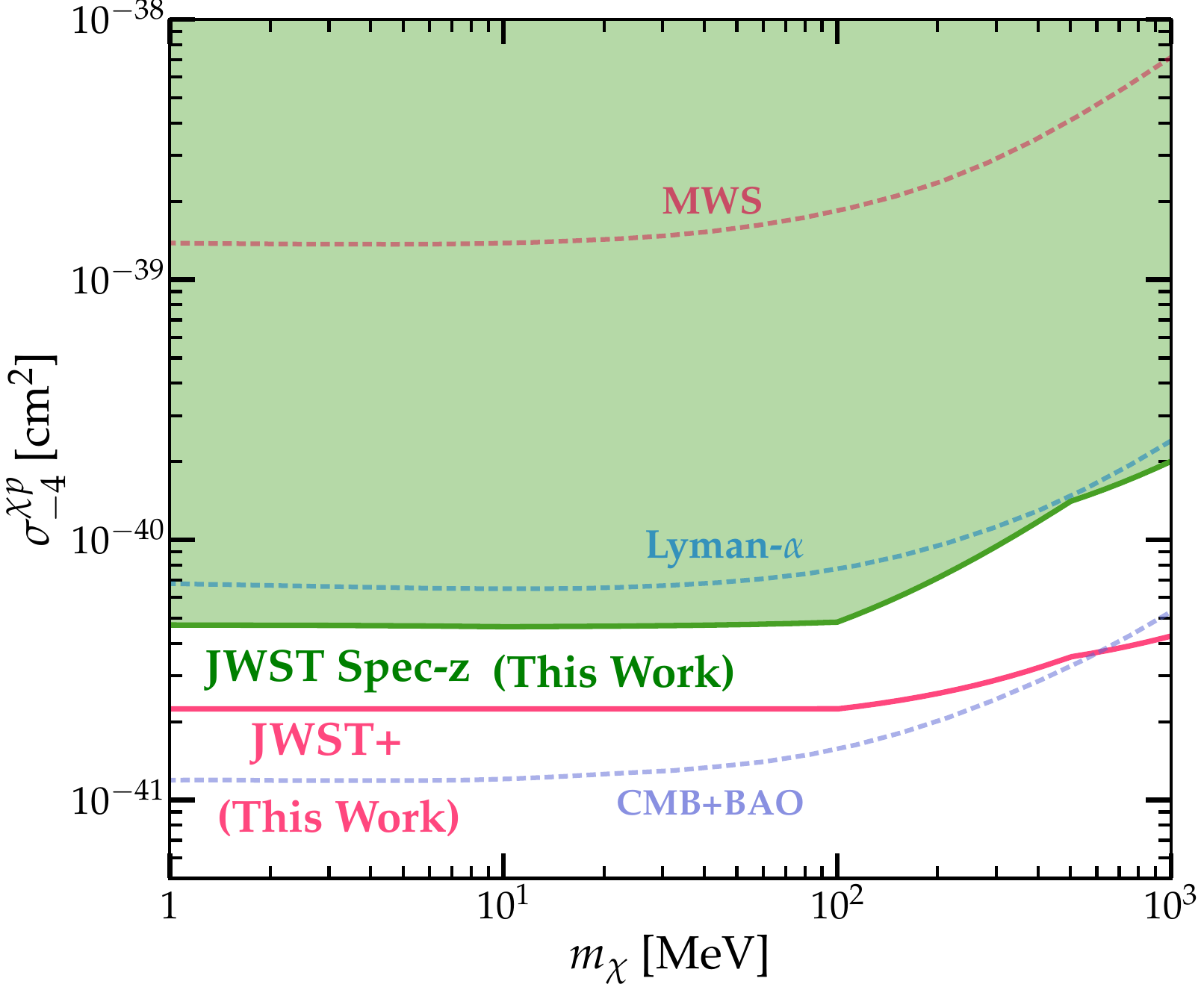}~~
		\includegraphics[width=\columnwidth]{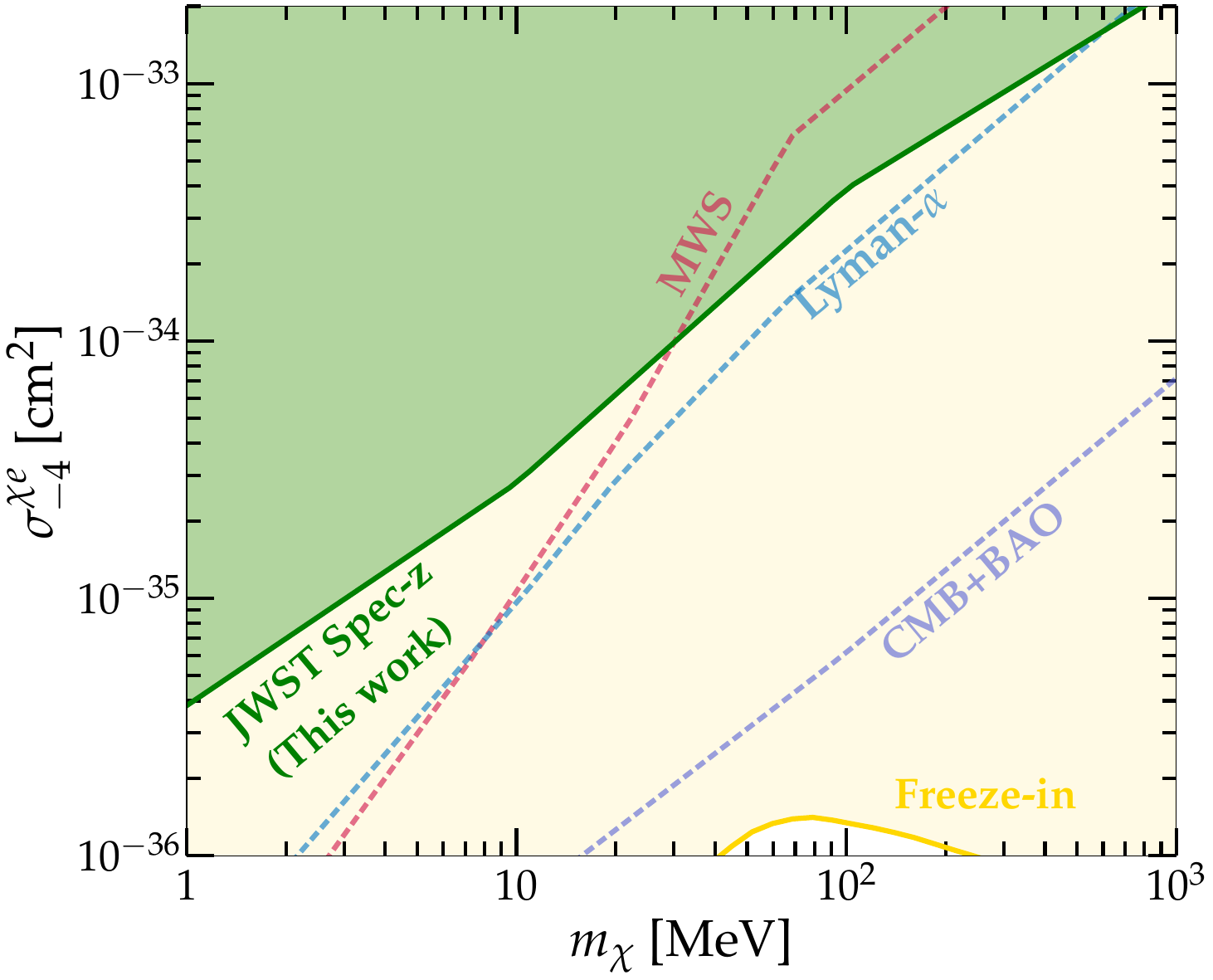}~~\\	
		\caption{95\% C.L.\,\,upper limits on the normalization of DM-baryon elastic scattering cross section for $n = -4$ case, as a function of the DM mass $m_\chi$. The \textbf{green} shaded region in both panels is ruled out by the ``JWST Spec-z" setup (our work). On the left panel, the \textbf{deep pink} solid line shows an upper limit on the \textit{Coulomb-like} interactions for ``JWST+" setup (our work). The various dashed curves represent other cosmological constraints -- Lyman-$\alpha$ forest (cyan), Milky Way satellite count (pink), and CMB+BAO (purple)\,\cite{Buen-Abad2022, Nguyen2021}. Some other constraints are shown in\,\cite{Buen-Abad2022}.
        \textbf{\textit{Left:}} Shows constraints for the DM-proton interactions. 
        \textbf{\textit{Right:}} Shows constraints for the DM-electron interactions. The yellow shaded region shows the constraint obtained assuming relic abundance from 100\% freeze-in\,\cite{Essig:2011nj, Chu:2011be}.
    }
		\label{fig:constraint-proton-electron-n-4}
	\end{center}	
\end{figure*}

Fig.~\ref{fig:uvlf-transfer-electron} shows the suppression in $T^{2}(k) = P(k)/P_{\Lambda\mathrm{CDM}}(k)$, which quantifies the suppression of the matter power spectrum, along with the UVLF for DM--electron interactions for different values of $n$. Here, $P(k)$ and $P_{\Lambda\mathrm{CDM}}(k)$ are the linear matter power spectra for IDM and $\Lambda$CDM cosmologies, respectively. For the interacting cases, $T^{2}(k)$ shows suppression at the smallest scales probed by JWST (shaded region on the left panel). The UVLF plot from the $\Lambda$CDM model provides a better fit to the data compared to the IDM cosmologies (right panel).

For the case with $n = -4$, this suppression is apparent starting at very large scales \cite{Buen-Abad2022}, while for $n = 0$ and $n = -2$ the suppression becomes significant at smaller scales. Owing to this feature, the strongest limits on \textit{Coulomb-like} $n=-4$ are derived from large scale observations such as the CMB and BAO\,\cite{Buen-Abad2022, Nguyen2021} while the other cases are best probed by observations at smaller scales such as Lyman$-\alpha$, Milky Way satellites, and JWST\,\cite{2024ApJ...976...40W}. However, due to the behaviour of the transfer function and JWST probing more early structures, ``JWST Spec-z" provides more stringent limits on \textit{Coulomb-like} interactions between DM and protons than other non-linear scale observables, such as Lyman-$\alpha$ and MWS. In the left panel of Fig.~\ref{fig:constraint-proton-electron-n-4}, as we used the ``JWST+" setup (deep pink solid line), the constraints on the normalization $\sigma^{\chi p}_{-4}$ have become stronger by a factor $\sim 2$ than the upper limits derived from ``JWST Spec-z" setup (green solid line).

\begin{figure}[h]
\centering
\includegraphics[width=\columnwidth]{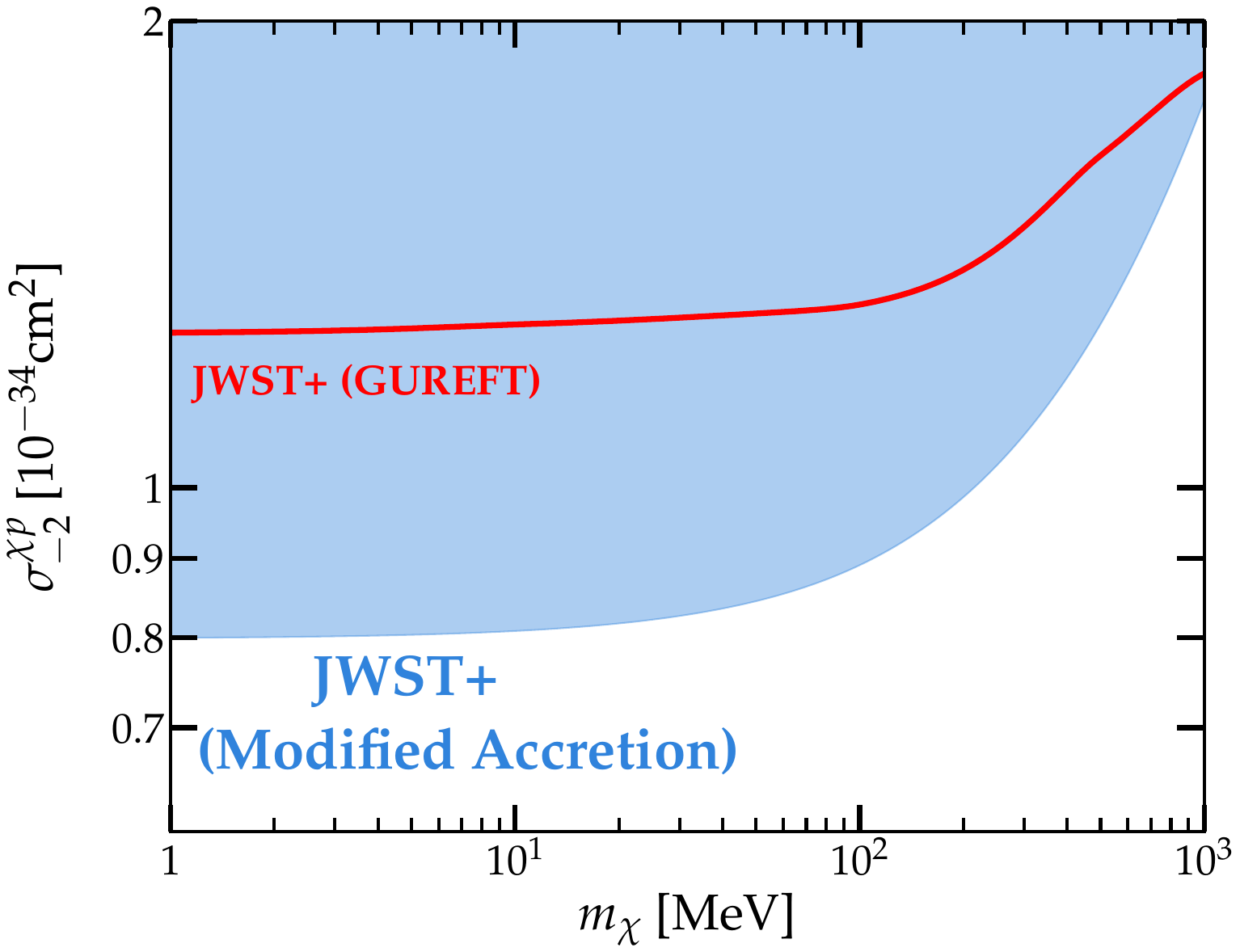}
\caption{Constraints on the normalization of the dark matter--proton scattering cross section
  $\sigma^{\chi p}_{-2}$ for \textit{electric-dipole-moment-like} $n=-2$ interactions are shown. The solid \textbf{red} line
  corresponds to constraints obtained using the empirical halo accretion rate, while the
  \textbf{blue} shaded region shows the constraints assuming the accretion rate is reduced by an
  order of magnitude at all redshifts and for all halo masses. Both the analysis is performed using the ``JWST+" setup.
}
\label{fig:Accretion}
\end{figure}

\begin{figure*}[htb]
\begin{flushleft}
\includegraphics[width=1.15\textwidth]{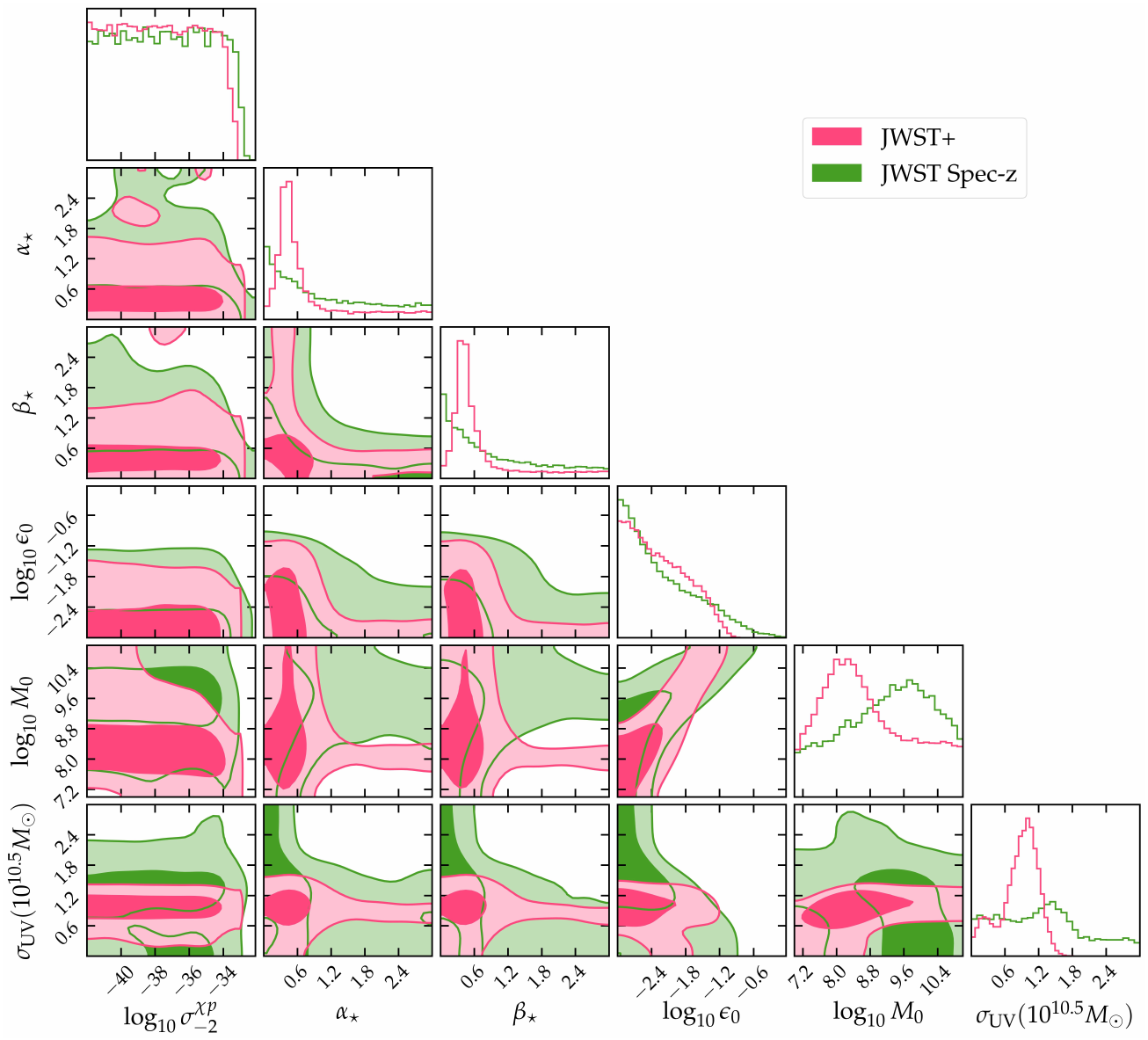}
\end{flushleft}
\caption{Corner plot of posterior distributions of the astrophysical parameters and $\sigma_{-2}^{\chi p}$ for $m_\chi = 1~\rm{MeV}$ with ``JWST Spec-z"(\textbf{green}) and ``JWST+" setup (\textbf{deep pink}). These values are in good agreement with best-fit values from ref.\,\cite{Shen:2025isu}. Since we include more data in the ``JWST+" setup, for this case most of the astrophysical parameters have better constrained 1D marginalized posteriors.
}
\label{fig:corner posteriors}
\end{figure*}

Fig.~\ref{fig:corner posteriors} shows the 1D and 2D marginalized posterior probability distributions for the \textit{electric-dipole-moment-like} $n = -2$ DM–proton interaction case with DM mass $m_{\chi} = 1\,\mathrm{MeV}$, for the cross-section $\sigma^{\chi p}_{-2}$ as well as the astrophysical model parameters $\{\alpha_{\star}, \beta_\star, \epsilon_0, M_0, \sigma_{\rm UV}(10^{10.5} M_\odot)\}$, for both the ``JWST Spec-z'' setup (in green) and the ``JWST+'' setup (deep pink).\,\,The constraints on the astrophysical parameters are in good agreement with the best-fit results from high-redshift \textsc{Thesan-Zoom} simulations \cite{Shen:2025isu}. These posterior samples from the ``JWST Spec-z" setup were used to obtain fig.~3 of the main text. Note that for the ``JWST+" setup, the constraints on the DM–proton interaction cross-section become stronger than those from the ``JWST Spec-z" setup, because it includes the ultra-high-redshift $z \sim 17$ and $z \sim 25$ UVLF datapoints. In addition, with the ``JWST+" setup, the astrophysical parameters have better-constrained marginalized posteriors, since we include more datapoints here as compared to the ``JWST Spec-z" setup.

\section{Boltzmann Equations for Dark Matter Interacting with Baryons}
 
In this section, we describe the formalism for the calculation of linear matter power spectrum $P(k)$ in the presence of interactions between DM and baryons. 
We mainly follow the conventions in refs.\,\cite{Ma1995, Dvorkin2014}. We work in the \textit{conformal Newtonian gauge} and modify the standard \textit{Boltzmann equations} by including elastic scattering between DM and baryons. This interaction leads to heat and momentum transfer between the two species.

For a large class of models, the momentum-transfer cross section between the interacting dark matter $\chi$ and the baryonic component $B \in \{ e, p \}$ (we assume that the different baryonic components, namely electrons and baryons, are tightly coupled via the electromagnetic force. This allows us to describe their properties such as temperature using a single set of variables, labeled with the common index $b$) reduces to:
\begin{equation}
    \sigma_{T, n}^{\chi B}(v)=\sigma^{\chi B}_{n}\, v^n \,,
\end{equation}
where $\sigma_{n}^{\chi B}$ is the velocity-independent normalization of the cross section, $v$ is the relative velocity between DM and baryons, and the index $n$ is determined by the nature of the interaction. DM coupled to baryons via a heavy mediator would result in $n=0$, while an ultralight mediator would lead to $n=-4$. Another case of interest is $n=-2$, which results from a dipole-like interaction.

We assume that (i) the DM consists of a single particle species, \textit{i.e.}, 100\% of the DM is interacting, (ii) both DM and baryons are non-relativistic during the relevant period of cosmological history, and (iii) both follow \textit{Maxwellian} velocity distributions \cite{Buen-Abad2022, Dvorkin2014, Chen2002}. From these assumptions, the \textit{conformal} momentum-exchange rate due to DM-baryon interactions \cite{Dvorkin2014} is:
\begin{equation}
    R_{\chi-b}=a\sum_{B} \frac{Y_B \rho_b}{m_\chi+m_B} \sigma_n^{\chi B}c_n u_B^{n+1}\,,
    \label{eq:R-chi-baryon-dmb}
\end{equation}
where $a$ is the scale factor, $Y_B \equiv \rho_B / \rho_b$ is the mass fraction of the baryon component $B$ (so $Y_p\approx 0.75$ and $Y_e \approx 4\times 10^{-4}$), $\rho_b$ is the net energy density of baryons,
$c_n = \frac{2^{(5+n)/2}}{3\sqrt{\pi}} \, \Gamma\!\left( 3 + \frac{n}{2} \right)$ is a numerical factor, and $m_{\chi}$, $m_{B}$ are the masses of the interacting DM particle and baryon respectively. The velocity dispersion $u_B$ is related to the DM and baryon temperatures as:
\begin{equation}
    u_B^2=\frac{T_b}{m_B}+\frac{T_\chi}{m_\chi}+\frac{\langle V^2_{\rm bulk}\rangle}{3},\end{equation}
where $T_{\chi}$, $T_{b}$ are the temperature of the DM and baryon fluids respectively, and $\langle V_{\rm bulk}^{2}\rangle$ is the mean squared relative bulk velocity between these two fluids.

In the presence of DM-baryon interactions, the standard Boltzmann equations are modified as:
\begin{itemize}
    \item \textbf{Continuity equations:} The continuity equations in this case are unchanged as compared to the equations for non-interacting CDM:
    \begin{align}
        \dot{\delta}_{\chi}&=-\theta_\chi+3\dot{\phi}\,,\\
        \dot{\delta}_{b}&=-\theta_b+3\dot{\phi}\,,
    \end{align}
    where $\delta_j = \delta \rho_j /\bar{\rho}_j$ is the density contrast, and $\theta_j = \partial_i v^i_{j}$ is the velocity divergence of the species $j\in\{\chi, b\}$. One of the two metric perturbations in the Newtonian gauge is denoted by $\phi$\,\cite{Ma1995}.

    \item \textbf{Euler equations:} The effect of DM-baryon interactions manifests as a drag term in the Euler equations:
\begin{align}
    \dot{\theta}_\chi&=-aH\theta_\chi+k^2 c_\chi^2\delta_\chi+R_{\chi-b}\qty(\theta_b-\theta_\chi)+k^2\psi\label{eq:theta_chi}\,,\\
    \dot{\theta}_b&=-aH\theta_b+k^2 c_b^2\delta_b+R_{b-\chi}\qty(\theta_\chi-\theta_b)\nonumber\\&\quad+R_{b-\gamma}\qty(\theta_\gamma-\theta_b)+k^2\psi \,,
    \label{eq:theta_b}
\end{align}
where $aH$ is the \textit{conformal Hubble parameter}, $\psi$ is a metric perturbation, and $k$ is the wavenumber of perturbations. The coupling strengths $R_{\chi-b}$ and $R_{b-\chi}$ (and $R_{b-\gamma}$ between baryons and photons due to electromagnetic interaction) are related by:
\begin{align}
    R_{b-\chi}&=\frac{\rho_\chi}{\rho_b}R_{\chi-b}\,,\\
    R_{b-\gamma}&=\frac{4 \rho_\gamma a n_e \sigma_T}{3 \rho_b}\,.
\end{align}
where $\rho_{\chi, b, \gamma}$ are DM, baryon and photon energy densities, $n_e$ is the electron number density, and $\sigma_T$ is the Thomson cross-section. The second terms on the R.H.S.\,\,of equations \ref{eq:theta_chi}-\ref{eq:theta_b} depend on the sound speeds $c_{\chi, b}$:
\begin{align}
    c_b^2&=\frac{k_B T_b}{\mu_b}\qty(1-\frac{1}{3}\dv{\ln T_b}{\ln a})\,,\\
    c_\chi^2&=\frac{k_B T_\chi}{m_\chi}\qty(1-\frac{1}{3}\dv{\ln T_\chi}{\ln a})\,,
    \label{eq:c_s-chi}
\end{align}
where $\mu_b$ is the mean baryon mass.

\item \textbf{Temperature evolution:} The evolution of DM and baryon temperatures with time is given by:
\begin{align}
    \dot{T}_\chi&=-2aHT_\chi+ 2R'_{\chi-b}\qty(T_b-T_\chi)\label{eq:temp-chi-dmb}\,,\\
    \dot{T}_b&=-2aHT_b+ 2R'_{b-\chi}\qty(T_\chi-T_b)\nonumber\\&\quad+ 2R'_{b-\gamma}\qty(T_\gamma-T_b) \,.
    \label{eq:temp-baryon-dmb}
\end{align}
We assume that the temperature of the photon bath is not affected by these interactions. The rate coefficients $R'$ of heat exchange are related to those of momentum exchange as:
\begin{align}
    R'_{\chi-b}&=\frac{m_\chi}{m_\chi+m_B}R_{\chi-b}\,,\\
    R'_{b-\chi}&=\frac{\mu_b}{m_\chi+m_B}R_{b-\chi}\,,\\
    R'_{b-\gamma}&=\frac{\mu_b}{m_e}R_{b-\gamma} \,.
\end{align}
\end{itemize}
The above set of equations are solved numerically using a version of the code {\tt CLASS} \cite{Lesgourges2011, Blas2011}, modified to include DM-proton and DM-electron elastic interactions \cite{Buen-Abad2022}. The primary result of this calculation is the linear \textit{matter power spectrum} $P(k) \sim \expval{\delta^2(k)}$. Numerical calculations reveal that $P(k)$ gets suppressed at characteristic length scales depending on the DM-baryon scattering cross-section and nature of the interaction, as presented in Fig. \ref{fig:uvlf-transfer-electron}.

\section{Calculation of Halo Mass Function}
Next, we describe the formalism for calculating the halo mass function (HMF) from the linear matter power spectrum. We use the extended \textit{Press-Schechter} formalism \cite{Press1974, Sheth:1999mn, Parashari2023} that encapsulates the physics of ellipsoidal halo collapse. The HMF obtained from this semi-analytical model is:
\begin{equation}
    \frac{dn}{d\log M_h}(z) = M_h \, \frac{\rho_{m,0}}{M_h^2} \, f\bigl(\sigma(M_h)\bigr) \, \left| \frac{d\ln \sigma}{d\ln M_h} \right| \label{eq:HMF}, 
\end{equation}
where $\rho_{m,0}$ is the comoving matter density, and $\sigma^2(R)$ is the variance of the smoothed density contrast within a sphere of radius $R$:
\begin{equation}
    \sigma^{2}(R) = \frac{1}{2\pi^{2}} \int_{0}^{\infty} \mathrm{d}k \, k^{2} \, W^{2}(kR) \, P(k),
\end{equation}
where $P(k)$ denotes the linear matter power spectrum, 
and $W(kR)$ is the Fourier transform of a window function, often called \textit{kernel} or \textit{filter} function.\,\,In our analysis, we choose the real-space \textit{top-hat} filter function (or the smooth-$k$ filter function for the ``JWST+" setup).

The function $f(\sigma)$ is called the \textit{fitting function}. There are several fitting functions derived from numerical simulations \cite{Murray2013}. We choose the \textit{Sheth-Tormen} fitting function \cite{Sheth:1999mn}:
\begin{equation}
    f(\sigma) = A \sqrt{\frac{2a}{\pi}} \left[ 1 + \left( \frac{\sigma^2}{a \delta_c^2} \right)^{p} \right] 
    \frac{\delta_c}{\sigma} \exp\left( -\frac{a \delta_c^2}{2\sigma^2} \right)
\end{equation}
with best-fit values $A=0.3222, a=0.707, p=0.3$, and $\delta_c=1.686$ is the critical value of the linear density contrast for a gravitational collapse.

\section{Galaxy-halo connection prescription}
We describe an empirical relation between the total mass $M_h$ of a DM halo and the star formation rate (SFR) in the galaxy hosted by it. We use this connection prescription in our work to obtain the SFR and ultimately the UV luminosity function UVLF. 

As baryons are converted into stars, it is a plausible assumption that more massive halos host more luminous galaxies than lighter halos. 
Given that the initial density perturbations in our Universe are adiabatic, the SFR is given by \cite{Shen2024, Shen:2025isu}:
\begin{equation}
    \text{SFR}=\epsilon_\star f_b \dot{M}_h \,,\label{eq:SFR}
\end{equation}
where $\dot{M}_{\rm {h}}$ is the halo mass accretion rate, $f_{b} = \Omega_{b}/\Omega_{m} \approx 0.16$ is the baryon mass fraction (which we assume is equal to the cosmological mass fraction of baryons), and $\epsilon_{\star}$ is the star formation efficiency (SFE). SFE represents the fraction of accreted baryonic matter that goes into forming stars, and is parametrized as described in the main paper.

The accretion rate $\dot{M}_h$ in Eq. \ref{eq:SFR} is calibrated using $N$-body simulations, and we use the calibrations inferred in ref. \cite{2024MNRAS.530.4868Y}.
\begin{eqnarray}
\frac{\dot{M}_h}{\rm{M_\odot~yr^{-1}}} &= \beta(z)(M_{h, 12}E(z))^{\alpha(z)} \nonumber \,\,,\\
\alpha(z) &= 0.858 + 1.554 a - 1.176 a^{2}\,\,, \nonumber\\
\log{\beta(z)}  &= 2.578-0.989 a -1.545 a^{2},   \nonumber
\label{eq:HaloAccRate}
\end{eqnarray}
where $E(z) = H(z)/H_{0}$, $M_{h,12} = M_h/(10^{12} M_{\odot})$, and $a = 1/(1+z)$ is the scale factor.

Finally, we want to relate the SFR to the luminosity of the galaxy. Simulations of stellar populations have inferred an empirical relation between the SFR and specific UV luminosity of star-forming galaxies \cite{2022PhRvD.105d3518S}:
\begin{equation}
    L_{\nu}^{\mathrm{UV}} = \frac{1}{\kappa_{\mathrm{UV}}}\,\mathrm{SFR} \, ,
\end{equation}
the proportionality constant $\kappa_{UV}$ is sensitive to the stellar initial mass function (IMF). In this work, we adopt the value $\kappa_{\rm UV}=0.72\times 10^{-28} \:\rm{M_\odot.yr^{-1}.erg^{-1}.s\textcolor{red}{.Hz}}$ corresponding to \textit{Chabrier} IMF, following refs. \cite{Madau2014, Shen2024}. The absolute magnitude and the luminosity are related to each other as:
\begin{equation}
    \log_{10}\Big(\frac{L_{\nu}^{\mathrm{UV}}}{\mathrm{erg\,s^{-1}}}\Big) = 0.4(51.63-M_{\mathrm{UV}})
\end{equation}

The magnitude $M_{\rm UV}(M_h)$ obtained above is the median UV magnitude for a galaxy hosted by a halo of mass $M_h$. However, due to the stochastic nature of the formation of galaxies, the relation $M_{UV}-M_h$ can have a scatter. We model this using a Gaussian kernel centered about the median $M_{\rm UV}^{\rm med}$ and with a width $\sigma_{\rm UV}$, as mentioned in the main paper:
\begin{equation}
    p(M_{\rm UV}|M_h) = \frac{1}{\sqrt{2\pi}\sigma_{\rm UV}} \exp \qty(-\frac{\qty(M_{\rm UV}- M_{\rm UV}^{\rm med})^2}{2 \sigma_{\rm UV}^2})\,.
\end{equation}

Moreover, as highlighted in the \textit{End Matter}, we take into account the dust-attenuation from ref. \cite{2016ApJ...833...72B}, which match with \textsc{Thesan-Zoom} simulations at high $z$.

\section{Accretion Rates for other DM models}
Merger trees play a central role in determining the accretion of halos. Since halo formation is sensitive to the matter power spectrum $P(k)$, in cosmologies where $P(k)$ deviates from the vanilla cold dark matter model, we naturally expect the mass accretion rate $\dot{M}_{h}$ to also deviate from empirical models calibrated using CDM-only simulations\,\cite{2024MNRAS.530.4868Y, Shen:2025isu}.

Since there is no such empirical form for the accretion rate in IDM cosmologies, we redo our analysis for $n = -2$, \textit{electric-dipole-moment-like} DM–proton interactions by multiplying our benchmark $\dot{M}_{h}$ from CDM simulations, \textsc{Gureft} \cite{2024MNRAS.530.4868Y}, by an overall normalization factor $f = 0.1$; i.e., the accretion rate is reduced to 10\% of its empirical value at all redshifts and for all halo masses. Although this is a minimalistic approach and a robust analysis will require full N-body simulations for these models, as done by \cite{2012MNRAS.420.2318L}, or the development of a self-consistent extended Press–Schechter (EPS) formalism calibrated against N-body simulations, as performed by \cite{2013MNRAS.428.1774B}, this gives us a quantitative estimate of how much the constraints would change if we change the $\dot{M}_{h}$. In general, we expect our constraints to strengthen if the accretion rate decreases, since this would suppress the UVLF and inhibit any further suppression from the momentum transfer due to DM–baryon interactions. This does not continue indefinitely for arbitrarily smaller values of $\dot{M}_{h}$, since for very small accretion rates high-redshift halo formation becomes highly suppressed. Thus, we lose sensitivity to the DM–baryon cross-section ranges that suppress the $k$ modes that collapse to form halos at those high redshifts.

The solid red line in Fig.~\ref{fig:Accretion} shows the constraints on the velocity--dependent DM-proton elastic scattering cross section for the $n=-2$ case, obtained using the empirical halo accretion model from \textsc{Gureft} simulations\,\cite{2024MNRAS.530.4868Y} in ``JWST+" setup. The blue shaded region is excluded by the ``JWST+" setup, with the accretion rate suppressed to $10\%$ of its fiducial value across all redshifts and halo masses. As expected, the limits become stronger in the latter case, since a lower accretion rate already suppresses the UV luminosity function, making any additional suppression from DM–proton interactions more strongly constrained.
\end{document}